\documentstyle[preprint,multicol,aps,prd,eqsecnum,amssymb,psfig]{revtex}

\newcommand{\beq} {\begin{equation}}
\newcommand{\eeq} {\end{equation}}
\newcommand{\beqa}{\begin{eqnarray}}
\newcommand{\eeqa}{\end{eqnarray}}

\newcommand{\dEdt}{\frac{{\mathrm d}E}{{\mathrm d}t}}
\newcommand{\rmd} {{\mathrm d}}

\newcommand {\AFF}  {de Alfaro, Fubini and Furlan}
\newcommand {\LS}   {L\"{u}\-scher-Schech\-ter}
\newcommand {\LC}   {Levi-Civita}
\newcommand {\YM}   {Yang--Mills}
\newcommand {\YMth} {Yang--Mills theory}

\newcommand {\YMHth}{Yang--Mills--Higgs theory}

\newcommand {\ew}   {electroweak}
\newcommand{\ewsm}  {electroweak Standard Model}
\newcommand {\sph}  {spha\-le\-ron}

\newcommand {\rhs}  {right-hand side}
\newcommand {\diffeq} {differential equation}
\newcommand {\diffeqs}{differential equations}
\newcommand {\zeieq}  {zero-eigenvalue equation}

\newcommand {\zenfeqs}{zero-energy fermion equations}
\newcommand {\tReq}   {transformed Riccati equa\-ti\-on}

\begin{document}
\draft
\tightenlines  \renewcommand{\baselinestretch}{1.4}

\title{Spectral flow of chiral fermions in nondissipative Yang--Mills
       gauge field backgrounds}

\author{F.R. Klinkhamer\thanks{E-mail: frans.klinkhamer@physik.uni-karlsruhe.de}
        and Y.J. Lee\thanks{E-mail:lee@particle.physik.uni-karlsruhe.de}}
\address{Institut f\"ur Theoretische Physik, Universit\"at Karlsruhe,
         D--76128 Karlsruhe, Germany}

\date{preprint  KA--TP--09--2001,  hep-th/0104096; Physical Review D 64, 065024 (2001)}

\maketitle
\begin{abstract}\small
Real-time anomalous fermion number violation is investigated for
massless chiral fermions  in spherically
symmetric $SU(2)$ Yang--Mills gauge field  backgrounds
which can be weakly dissipative or even nondissipative.
Restricting consideration to spherically symmetric fermion fields,
the zero-eigenvalue equation of the time-dependent effective Dirac
Hamiltonian is studied in detail. For generic spherically symmetric $SU(2)$
gauge fields in Minkowski spacetime, a relation is presented
between the spectral flow and
two characteristics of the background gauge field. These characteristics
are the well-known ``winding factor,'' which is defined to be the change
of the Chern-Simons number of the associated vacuum sector of the background
gauge field, and a new ``twist factor,'' which can be obtained from the
zero-eigenvalue equation of the effective Dirac Hamiltonian but is entirely
determined by the background gauge field.
For a particular class of  (weakly dissipative) L\"uscher-Schechter
gauge field solutions, the level crossings are calculated directly and
nontrivial contributions to the spectral flow from both
the winding factor and  the twist factor are observed.
The general result for the spectral flow may be
relevant to electroweak baryon number violation in the early universe.
\end{abstract}

\section{Introduction}

An important consequence of the triangle anomaly \cite{A69,BJ69} is the
violation of fermion number conservation in the \ewsm~if the
background gauge field has nonvanishing topological charge $Q$.
This connection between the triangle anomaly and fermion number
violation in the electroweak Standard Model was first pointed out in Ref.
\cite{H76}. For the background gauge field, the calculation of Ref. \cite{H76}
used Euclidean instanton solutions \cite{BPST75} in order to
calculate the tunneling amplitude between topologically different vacua.
A special feature of the Euclidean (imaginary-time) approach is that
gauge field configurations with finite action fall into homotopy classes
labeled by an \emph{integer} $Q$. This integer $Q$ then gives
the number of fermions produced.

For real-time fermion-number-violating processes \cite{C80}, sphaleron-like
gauge field configurations are believed to play crucial role \cite{KM84a}.
These gauge fields ``interpolate'' between topologically different vacua
and have sufficiently high energy to overcome the energy barrier.
But, in contrast with the Euclidean approach, the anomalous fermion production in
Minkowski spacetime is not, in general, given by the topological
charge $Q$ of the classical gauge field background.
The reason is that $Q$ may be a noninteger or even, for the case of \YMHth,
not well-defined; cf. Refs. \cite{FKS93,FGGRS95}.
It is not clear which quantity, in general, determines the anomalous
fermion production for real-time processes.

For pure \YMth~in Minkowski spacetime, the authors of Refs.
\cite{GH95,K95} have argued that the number of produced fermions
is given by the change of winding number of the
associated vacua of the initial and final gauge field configurations.
The ``associated vacua'' of a given classical gauge field background represent
the particular vacuum
configurations that the background field would approach as $t\rightarrow
-\infty$ or $t\rightarrow +\infty$. As is clear from the context, the
authors of Refs. \cite{GH95,K95} considered  \emph{dissipative} background
fields in order to be able to quantize the fermion fields at $t=\pm\infty$.

The question, now, is what happens to fermion number violation if
a powerful energy source creates a nontrivial gauge field background over an
\emph{extended}  spacetime region (for example, in a high-energy collision
experiment or in the early universe). In this case, we cannot readily associate
the initial high-energy state with a particular vacuum configuration and
the field-theoretic approach used in Refs. \cite{GH95,K95} breaks down.
For nontrivial classical bosonic background fields,
it is, moreover, not known how to construct a fermion number operator in
terms of the quantized fermionic fields .

Still, fermion number violation can be directly observed from the level
crossing of the energy eigenvalues of the time-dependent Dirac Hamiltonian,
see Ref. \cite{C80} and references therein. The
overall effect of level crossing can be characterized by the ``spectral
flow,'' defined to be the number of eigenvalues of the Dirac
Hamiltonian that cross zero from below minus the number of eigenvalues that cross
zero from above, for a given time interval and direction of time.

In this paper, we study the zero-eigenvalue equation of the effective
Dirac Hamiltonian for spherically
symmetric chiral fermion fields and
classical $SU(2)$ Yang--Mills gauge field backgrounds. A relation is found
between the  spectral flow and certain features of the
spherically symmetric background gauge field.
These features are the well-known topological ``winding factor'' and a
new type of ``twist factor,'' both of which will be defined later. The spherically
symmetric subspace of (3+1)-dimensional chiral $SU(2)$ Yang--Mills theory is
equivalent to an (1+1)-dimensional $U(1)$ gauge theory coupled to a Higgs-like
complex scalar field \cite{W77} and several two-component Dirac fields.
This drastic simplification allows us to examine the problem using analytical
methods.

The (3+1)-dimensional spherically symmetric $SU(2)$ gauge field backgrounds
considered in this paper are, in general,
\emph{nondissipative}, which means that the energy density
does not approach zero uniformly as $t$ $\rightarrow$ $\pm\infty$.
For these nondissipative gauge field backgrounds, the
contribution of the twist factor to the spectral flow is manifest.
The general result for the  spectral flow also applies to dissipative
spherically symmetric gauge
field backgrounds, which will be classified later as ``weakly dissipative''
and ``strongly dissipative.'' It will be shown
that the nonvanishing effect of the twist factor for the spectral flow
can already appear for \emph{weakly dissipative} spherically symmetric
gauge field backgrounds, such as certain
\LS~gauge field solutions \cite{L77,S77}.
Our paper may, therefore, be
viewed as a continuation of the work of Refs. \cite{C80,GH95,K95}.

The outline of this paper is as follows. In Section II, we present the
model and the basic formalism. After giving the chiral $SU(2)$
Yang--Mills theory in the spherically symmetric \emph{Ansatz}, we briefly
review the topological properties of the gauge field background.
The gauge field topology is, in the first place,
characterized by the winding factor,
defined to be the change of the Chern-Simons number of the
associated vacuum sector of the gauge field configuration.

In Section III, we
consider the zero-eigenvalue equation of the time-dependent effective Dirac
Hamiltonian. By investigating the zero-eigenvalue equation directly, we
are able to identify a family of Riccati equations \cite{I56,H69}, from which
the twist factor of the spherically symmetric $SU(2)$
gauge field configuration can be obtained.

In Section IV, we present a result
for generic spherically symmetric $SU(2)$ gauge field backgrounds,
which relates the spectral flow to both the winding factor and the twist factor.

In Section V, we investigate the level crossing phenomenon for the particular
spherically symmetric $SU(2)$ gauge field backgrounds given by certain
\LS~solutions \cite{L77,S77}
and verify our relation for  the spectral flow. Specifically, we demonstrate
the significant effect of the twist factor for a
class of \LS~background gauge fields with energies far above a \sph-like barrier.

In Section VI,
finally, we summarize our results and briefly discuss the role of dissipation.
There is also an appendix, which provides the proof of a result needed in this section.

For the benefit of the reader, we remark that
Sections II B, III C, and IV B form the core of the paper.

\section{Chiral $SU(2)$ Yang--Mills theory}

In this section, we review the spherically symmetric \emph{Ansatz} for massless chiral
fermions coupled to classical $SU(2)$ Yang--Mills gauge fields and establish
our notation. Furthermore, we recall the definition of the topological winding factor.

\subsection{Spherically symmetric \emph{Ansatz}}

The $SU(2)$ Yang--Mills theory with massless chiral fermions is described
by the action,
\begin{eqnarray}\label{action4d}
S &=& S_{G}+S_{F},\quad
S_{G} = -\frac{1}{2\,g^{2}}\int_{{\mathbb{R}}^4} {\mathrm d}^{4}x \: \mbox{Tr}
\left(F^{mn}F_{mn}\right),\quad
S_{F}=\int_{{\mathbb{R}}^4} {\mathrm d}^{4}x\:\bar{\Psi}_f\,\Gamma^{m}D_{m}\Psi_f ,
\label{SF-4d}
\end{eqnarray}
where $S_{G}$ represents the gauge field action and $S_{F}$ the fermionic
action.
Latin indices $m$, $n$, etc. run over the coordinate labels 0, 1,
2, 3, and the metric tensor for flat Minkowski spacetime is
$\eta_{mn}=\mbox{diag}(-1,1,1,1)$. Repeated indices are summed over.
The flavor index $f$, in particular, is summed over $1$, $\ldots$ , $N_F$.
Also, natural units are used for which $c$ $=$ $\hbar$ $=$ $1$.

The $SU(2)$ field strength tensor $F_{mn}$ and the covariant
derivative $D_{m}$ for the fermionic fields are defined as follows:
\begin{eqnarray}
F_{mn}&\equiv&\partial_{m}A_{n}-\partial_{n}A_{m}
+[A_{m},A_{n}], \quad A_{m}\equiv
A_{m}^{a}\,\tau^{a}/(2i)\;, \nonumber\\
D_{m} &\equiv& \partial_{m}+A_{m}\,\mbox{P}_{L}, \quad
\mbox{P}_{L}\equiv \left(1-\Gamma_{5}\right)/2, \quad
\mbox{P}_{R}\equiv \left(1+\Gamma_{5}\right)/2\;. \label{notation-4d}
\end{eqnarray}
The Dirac matrices $\Gamma^m$ are taken in the chiral (Weyl)
representation,
\begin{eqnarray}
\Gamma^{0} &=& -i\left(\begin{array}{cc}
0 & \openone\\
\openone & 0
\end{array}\right), \quad \Gamma^{a}=-i\left(\begin{array}{cc}
0 & \sigma^{a}\\
-\sigma^{a} & 0
\end{array}\right), \quad
\Gamma_{5}\equiv-i\,\Gamma^{0}\Gamma^{1}\Gamma^{2}\Gamma^{3}=
\left(\begin{array}{cc}
\openone & 0\\
0 & -\openone
\end{array}\right),\nonumber\\[0.2cm]
\openone&\equiv& \left( \begin{array}{cc} 1 & 0 \\ 0 & 1\end{array}\right), \quad
\sigma^1 \equiv \left( \begin{array}{cc} 0 & 1 \\ 1 & 0 \end{array}\right) , \quad
\sigma^2 \equiv \left( \begin{array}{cc} 0 & -i \\ i & 0 \end{array}\right), \quad
\sigma^3 \equiv \left( \begin{array}{cc} 1 & 0 \\ 0 & -1 \end{array}\right)  \mbox{.}
\label{gamma-4d}
\end{eqnarray}
The conjugate spinor is given by
$\bar{\Psi}_f$ $\equiv$ $\Psi_f^{\dagger}(-i\,\Gamma^{0})$.
Here and in the following, $\tau^{a}$ and $\sigma^{a}$ are Pauli matrices
carrying isospin and spin indices, respectively. The action (\ref{action4d})
thus corresponds to a chiral $SU(2)$ gauge theory, with interacting
left-handed fermions ($\Psi_{Lf}$ $\equiv$ $P_L\Psi_f$) and
noninteracting right-handed fermions ($\Psi_{Rf}$ $\equiv$ $P_R\Psi_f$).

The total number $N_F$ of flavors in the fermionic action (\ref{action4d})
must be even, in order to cancel the nonperturbative $SU(2)$ anomaly \cite{W82}.
Henceforth, we focus on a single flavor and drop the index $f$.
Since there is no natural mass
scale for the classical $SU(2)$ Yang--Mills theory, we also take an arbitrary
mass scale to work with. (In the full theory, quantum effects may, of course, fix the
scale.)

In this paper, we concentrate on the spherically symmetric subspace of the
(3+1)-dimensional theory. We use the following \emph{Ansatz} for the gauge
fields:
\begin{mathletters} \label{Ansatz-gauge}
\beqa
A_{0}(x)&=&\frac{1}{2i}\,a_{0}(t,r)\,\vec{\tau}\cdot\hat{\mbox{\bf x}}\;,\\[0.2cm]
A_{a}(x)&=&\frac{1}{2i}\left[\,\frac{\alpha(t,r)-1}{r}\,
\epsilon_{abc}\hat{x}_{c}\tau_{b}+\frac{\beta(t,r)}{r}\,
(\delta_{ab}-\hat{x}_{a}\hat{x}_{b})\tau_{b}+a_{1}(t,r)\,
(\vec{\tau}\cdot\hat{\mbox{\bf x}})\hat{x}_{a}\right],
\eeqa
\end{mathletters}
where $\alpha$, $\beta$, $a_0$, and $a_1$ are
\emph{real} functions of $t$ and $r$; cf. Refs. \cite{W77,Y89}. These gauge fields are
invariant under spatial $SO(3)$ rotations, up to a gauge transformation
\beq \label{gaugetransf}
A_{m}\rightarrow A_{m}^{\Omega}\equiv
                               \Omega\, (\partial_{m}+A_{m})\, \Omega^{\dagger},
\eeq
with $\Omega (x)\in SU(2)$.

The spherically symmetric \emph{Ansatz} for the fermionic fields is given by
(see Ref. \cite{Y89} and references therein)
\begin{mathletters} \label{Ansatz-fermion}
\begin{eqnarray}
\Psi(x) &=&\left(\begin{array}{c}
\Psi_{R}(x) \\
\Psi_{L}(x)
\end{array}\right)\equiv \left(\begin{array}{c}
\tau^{2}\,\tilde{\Psi}_{R}(x) \\
\tilde{\Psi}_{L}(x)
\end{array}\right),\\
\tilde{\Psi}_{L}(x)&=&\frac{1}{\sqrt{2}}\left[H_{L}(t,r)^{\phantom{\,}}+^{\phantom{*}}
i\,G_{L}(t,r)\,\vec{\tau}\cdot\hat{\mbox{\bf x}}\right]
\left(\begin{array}{c}
\left(\begin{array}{c}
0\\
+1
\end{array}\right)_{\mathrm{isospin}}\\
\left(\begin{array}{c}
-1\\
0
\end{array}\right)_{\mathrm{isospin}}
\end{array}\right)_{\mathrm{spin}},\\
\tilde{\Psi}_{R}(x) &=&\frac{1}{\sqrt{2}}\left[H_{R}(t,r)+
i\,G_{R}(t,r)\,\vec{\tau}^{\,*}\cdot\hat{\mbox{\bf x}}\right]
\left(\begin{array}{c}
\left(\begin{array}{c}
+1\\
0
\end{array}\right)_{\mathrm{isospin}}\\
\left(\begin{array}{c}
0\\
+1
\end{array}\right)_{\mathrm{isospin}}
\end{array}\right)_{\mathrm{spin}},
\end{eqnarray}
\end{mathletters}
where $H_{L}$, $H_{R}$, $G_{L}$, and $G_{R}$ are \emph{complex} functions
of $t$ and $r$. In components ($a$ for isospin and $\alpha$ for spin),
the two constant spinors of Eqs. (\ref{Ansatz-fermion}b,c)
can be written  as
$\epsilon_{a\alpha}$ and $\delta_{a\alpha}$, respectively,
where $\epsilon$ and $\delta$ are the \LC~and Kronecker symbols.

Furthermore, we assume that all physical (3+1)-dimensional field configurations
are described by infinitely differentiable functions
(this assumption can be relaxed). In order to have
regular behavior at the spatial origin for the (3+1)-dimensional field configurations
and their derivatives, the \emph{Ansatz} functions should satisfy the
following $r$-parity expansions near $r=0$:
\beq
\begin{array}{rclcrcl}
a_{0}(t,r)  &=&\sum_{k=0}^{\infty}a_{0}^{(2k+1)}(t)\, r^{2k+1},  &&
a_{1}(t,r)  &=& \sum_{k=0}^{\infty}a_{1}^{(2k)}(t)\, r^{2k},
\\[0.2cm]
\alpha (t,r)&=&1+\sum_{k=1}^{\infty}\alpha^{(2k)}(t)\, r^{2k}, &&
\beta (t,r) &=&\sum_{k=0}^{\infty}\beta^{(2k+1)}(t)\, r^{2k+1},
\;\; \beta^{(1)}(t) =  a_{1}^{(0)}(t),\\[0.2cm]
H_{L,R}(t,r)&=&\sum_{k=0}^{\infty}H_{L,R}^{(2k)}(t)\, r^{2k},  &&
G_{L,R}(t,r)&=&\sum_{k=0}^{\infty}G_{L,R}^{(2k+1)}(t)\, r^{2k+1},
\label{regular}
\end{array}
\eeq
with the expansion coefficients depending on time only.

If we substitute the \emph{Ans\"atze} (\ref{Ansatz-gauge}) and
(\ref{Ansatz-fermion}) into
the action (\ref{action4d}), the following reduced actions are obtained
(for a single fermion flavor):
\begin{mathletters} \label{S-2d}
\begin{eqnarray}
S_{G}&=&\frac{4\pi}{g^2}\int_{-\infty}^{+\infty}
{\mathrm d}t\int_{0}^{\infty} {\mathrm d}r
\left\{\frac{1}{4}\,r^2f_{\mu\nu}f^{\mu\nu}+|D_{\mu}\chi
|^{2}+\frac{1}{2\,r^2}(|\,\chi
|^{2}-1)^{2}\right\},\label{SG-2d}\\[0.2cm]
S_{F}&=&4\pi\int_{-\infty}^{+\infty}
{\mathrm d}t\int_{0}^{\infty} {\mathrm d}r
\left\{\bar{\Psi}_{l}\left(\gamma^{\mu}D_{\mu}
+\frac{1}{r}(\mbox{Re}\chi+i\gamma_{5}\,\mbox{Im}\chi
)\right)\Psi_{l}
+\bar{\Psi}_{r}\left(\gamma^{\mu}\partial_{\mu}
+\frac{1}{r}\right)\Psi_{r}\right\}.\nonumber\\
& &\label{SF-2d}
\end{eqnarray}
\end{mathletters}
Greek indices $\mu$, $\nu$, etc. run over the coordinate labels 0, 1, and are
lowered with the metric $\eta_{\mu\nu}\equiv \mbox{diag}(-1,1)$. The
coordinates $(x^0,x^1)$ correspond to $(t,r)$.
The theory (\ref{S-2d}) can be interpreted as an (1+1)-dimensional $U(1)$ gauge field
theory with a Higgs-like complex scalar
field $\chi(t,r)$ and two-component Dirac spinors $\Psi_l(t,r)$ and $\Psi_r(t,r)$.
In terms of the \emph{Ansatz} functions,
the $U(1)$ field strength $f_{\mu\nu}$,
the complex scalar and Dirac fields, and the covariant derivatives are given by:
\begin{eqnarray}
f_{\mu\nu}&\equiv&\partial_{\mu}a_{\nu}-\partial_{\nu}a_{\mu},\quad
\chi \equiv\alpha  +i\beta , \quad
D_{\mu}\chi \equiv(\partial_{\mu}-ia_{\mu})\chi ,\nonumber\\[0.2cm]
\Psi_{l}(t,r)&\equiv&\left(\begin{array}{c}
                           \Psi_{l1}(t,r) \\ \Psi_{l2}(t,r) \end{array}\right)
              \equiv \left(\begin{array}{c}
                           r\,H_{L}(t,r)\\ r\,G_{L}(t,r) \end{array}\right)
, \quad
 \Psi_{r}(t,r)\equiv\left(\begin{array}{c}
r\,H_{R}(t,r)\\
r\,G_{R}(t,r) \end{array}\right), \nonumber\\[0.2cm]
\bar{\Psi}_{l,r}&\equiv&\Psi_{l,r}^{\dagger}(-i\gamma^{0}),\quad
D_{\mu}\Psi_l
  \equiv\left(\partial_{\mu}+i\,(a_{\mu}/2)\,\gamma_{5}\right)\Psi_l \mbox{,}
\label{notation-2d}
\end{eqnarray}
with
\begin{eqnarray}
& & \gamma^{0}=i\sigma^{1}, \quad \gamma^{1}=-\sigma^{3}, \quad
    \gamma_{5}=-\gamma^{0}\gamma^{1}=\sigma^{2}\mbox{.}
\label{gamma-2d}
\end{eqnarray}

The spherically symmetric \emph{Ansatz} (\ref{Ansatz-gauge}),
(\ref{Ansatz-fermion}) preserves a $U(1)$ subgroup of the
$SU(2)$ gauge group, with transformation parameters $\Omega (\mbox{\bf
x},t)=\exp [\,i\,\omega (t,r)\,\tau\cdot\hat{\mbox{\bf x}}\,/\,2\,]$
in Eq. (\ref{gaugetransf}) above. Under these particular $SU(2)$ gauge
transformations, we have for
the (1+1)-dimensional fields the following $U(1)$ gauge transformations:
\beq
a_{\mu}\rightarrow a_{\mu}+\partial_{\mu}\omega , \quad
\chi \rightarrow e^{i\omega}\chi , \quad  \Psi_{l}\rightarrow
e^{-i (\omega/2)\gamma_{5}}\,\Psi_{l},
\quad \Psi_{r}\rightarrow \Psi_{r}\mbox{.}
\label{gaugeT-2d}
\eeq
In order to maintain the regularity of the (3+1)-dimensional field configurations,
$\omega (t,r)$ should have an odd $r$-parity expansion near $r=0$,
\beq
\omega (t,r)=\sum_{k=0}^{\infty}\omega^{(2k+1)}(t)\,r^{2k+1},
\label{regular-omega}
\eeq
where the expansion coefficients are dependent only on time.

For later reference, the (1+1)-dimensional
fields with finite energy approach a vacuum configuration at infinity, provided
\beq
\chi\rightarrow e^{i\omega},\quad
D_{\mu}\chi\rightarrow 0,\quad
a_{\mu}\rightarrow\partial_{\mu}\,\omega ,\quad
f_{\mu\nu}\rightarrow 0,\quad
\Psi_{l,r} \rightarrow 0,\quad {\mathrm for}\quad r \rightarrow \infty\;,
 \label{r-infinity}
\eeq
and are regular at the spatial origin, provided
\beq
|\,\chi| \rightarrow 1,
\quad D_{\mu}\,\chi\rightarrow 0,
\quad \Psi_{l,r} \rightarrow 0 ,\quad {\mathrm for}\quad r \rightarrow 0\;.
\label{r-zero}
\eeq
See Ref. \cite{Y89} for further details. Throughout this paper, we consider
regular spherically symmetric $SU(2)$ gauge fields with finite energy.

\subsection{Gauge field winding factor}

For the description of the topology of spherically symmetric $SU(2)$
gauge field backgrounds, it is convenient
to express the (1+1)-dimensional complex field $\chi(t,r)$ in polar form:
\beq
\chi (t,r)=\rho (t,r)\exp\left[\,i\,\varphi (t,r)\right], \quad
\rho (t,r) \geq 0 \mbox{.}\label{chi-polar}
\eeq
The ``associated vacuum sector'' of the background gauge field at a fixed time $t$ is
obtained from the configuration with $\rho(t,r)$ replaced
by $1$, but with $\varphi(t,r)$ and $a_\mu(t,r)$ unchanged.
[Note that the resulting configuration with $\rho(t,r)=1$ may still have nonzero energy
density (\ref{SG-2d}).]
For the gauge choice $\chi (t,0)=\chi (t,\infty )=1$, the
integer winding number is then defined as
\beq
N_{\chi}(t)\equiv [\,\varphi (t,\infty )-\varphi(t,0)\,]/(2\pi). \label{windingN}
\eeq
This winding number $N_{\chi}(t)$ is, in fact,
equal to the Chern-Simons number of the $\rho=1$ gauge field at time
$t$; see  Eq. (2.18) of Ref. \cite{Y89}. For a particular $\chi(t,r)$
configuration, the winding number $N_{\chi}(t)$ is unambiguous, provided
$|\chi(t,r)|$ $>$ $0$. See Sec. IV B for further discussion.

For a time interval $[\,t_i,t_f\,]$ with $t_i < t_f$, generic spherically
symmetric $SU(2)$ gauge field backgrounds
are characterized by the change of winding number $N_{\chi}$ between the
initial and final configurations,
\beq
\Delta N_{\chi}[\,t_f,t_i\,]\equiv
N_{\chi}(t_{f})-N_{\chi}(t_{i})\mbox{.}\label{windingF}
\eeq
Henceforth, we call $\Delta N_{\chi}[\,t_f,t_i\,]$ as defined by Eq. (\ref{windingF})
the ``winding factor'' of the spherically symmetric $SU(2)$ gauge field.
Our definition of the ``winding factor'' is directly inspired by the results
of Ref. \cite{K95}, obtained for a particular class of background fields
that will be discussed further in Sec. V.

\section{Time-dependent Dirac Hamiltonian and twist factor}

In this section, we consider the zero-eigenvalue equation of the time-dependent
effective Dirac Hamiltonian for a given
spherically symmetric $SU(2)$ gauge field configuration at one particular time.
The existence of fermion zero modes  is discussed and a necessary condition derived.
In addition, the so-called twist factor is introduced, which will play an
important role in Sec. IV.

\subsection{Fermion zero modes and level crossings}

The general solution $\Psi_l(t,r)$ of the (1+1)-dimensional Dirac equation from the
action
(\ref{SF-2d}) can be expressed as a linear combination of the eigenfunctions of the
corresponding time-dependent Dirac  Hamiltonian.
The eigenvalue equation of this Hamilton operator is
\begin{mathletters}
\begin{eqnarray}
H(t,r)\, \Psi (t,r)&=&E(t)\,\Psi (t,r), \label{Hamiltonianeq}\\[0.2cm]
H(t,r)             &\equiv& \gamma_{5}\,a_{0}/2 -i \gamma_{5}\,D_{1} \nonumber\\
                   &      & +\,i \gamma^{0} \left(\mbox{Re}\chi
                            +i\,\mbox{Im}\chi\,\gamma_{5}\right)/r,\label{Hamiltonian}
\end{eqnarray}
\end{mathletters}
where the covariant derivative $D_1$ has been defined in Eq. (\ref{notation-2d})
and $\Psi$ now stands for the two-component Dirac spinor $\Psi_{l}$ of
that same equation. [The other Dirac field
$\Psi_r(t,r)$ of the action (\ref{SF-2d}) has no interactions and will not be
considered in the following.]
The Hamiltonian (\ref{Hamiltonian}) depends on $t$ and $r$ through the background
fields $\chi(t,r)$ and $a_\mu(t,r)$, together with an explicit dependence
on $r$ in the $i \gamma^0$ term.

It is known that the zero-crossing of an energy eigenvalue of the Dirac Hamiltonian
is one of the crucial ingredients of fermion number violation; cf. Refs. \cite{C80,G94}.
In our case, the zero-eigenvalue equation
(\ref{Hamiltonianeq}) at fixed time $t$ can be written as
\begin{mathletters} \label{FZMeq-ND}
\begin{eqnarray}
\partial_r \Psi &=&{\cal A}\, \Psi , \quad {\cal A}\equiv {\cal
A}_{H}+{\cal A}_{A}, \label{FZMeq-ND-A}\\
{\cal A}_{H}&\equiv& - \left(\gamma^{1}\,\mbox{Re}\chi+
i\gamma^{0}\,\mbox{Im}\chi\right)/r, \label{FZMeq-ND-AH}\\
{\cal A}_{A}&\equiv& -i\,a_{0}/2-i\gamma_{5} \,a_{1}/2 \mbox{,} \label{FZMeq-ND-AA}
\end{eqnarray}
\end{mathletters}
where $\partial_r$ stands for the partial derivative with respect to $r$.
For later convenience, we have decomposed the matrix
${\cal A}$ of Eq. (\ref{FZMeq-ND-A}) into a Hermitian part
${\cal A}_{H}$ and an anti-Hermitian part ${\cal A}_{A}$.
Recall that we use two-dimensional Dirac matrices $\gamma^{0}=
i\sigma^{1}$, $\gamma^{1}=-\sigma^{3}$, and $\gamma_{5}=\sigma^{2}$, with
$\sigma^{a}$ the standard $2\times 2$ Pauli matrices.

In order to have a
regular (3+1)-dimensional fermionic field at $r=0$, the (1+1)-dimensional
fermionic field $\Psi (t,r)$ must satisfy the boundary condition $\Psi
(t,0)= 0$, which is already implemented by the \emph{Ansatz} (\ref{notation-2d}).
A fermion zero mode is then defined to be a {\em normalizable} solution
of Eq. (\ref{FZMeq-ND-A}) with boundary condition $\Psi (t,0)=0$.
Specifically, the normalization condition is given by
\beq \label{norm}
\int_{0}^{\infty}{\mathrm d}r \: |\Psi (t,r)|^2 =1 \mbox{.}
\eeq

The existence of a fermion zero mode at a particular time $t=t^{*}$ does
not necessarily imply level crossing of the eigenvalue of the
Dirac Hamilton operator. In fact, the energy level $E(t)$
could just ``touch" the $E=0$ value instead of ``crossing" it. Therefore, it is
necessary to check that level crossing really occurs.
This can be done by calculating   the time-gradient of the energy
eigenvalue $E(t)$ at $t=t^{*}$. If ${\mathrm d}E /{\mathrm d}t
|_{t=t^{*}}\neq 0$, then there is level crossing at $t=t^{*}$.

The overall effect of level crossings can be characterized by the ``spectral
flow'' ${\cal F}[\,t_f,t_i\,]$, defined to be the number of
eigenvalues of the Dirac Hamiltonian $H(t,r)$ that cross zero from below
minus the number of eigenvalues that cross zero from above, for the time
interval $[\,t_i,t_f\,]$ considered. The spectral flow will be discussed further
in Sec. IV. Here, we continue the investigation of the zero-eigenvalue
equation \emph{per se}.

\subsection{Gauge-invariant zero-eigenvalue equation}

We first express the zero-eigenvalue equation (\ref{FZMeq-ND-A}) at a
fixed time $t$ in terms of a set of bosonic background fields that are
invariant under the $U(1)$ gauge transformations (\ref{gaugeT-2d}). As can be seen
from Eq. (\ref{FZMeq-ND}), a nonvanishing gauge field $a_{0}$ contributes only
a complex phase factor to the solution $\Psi$. Up to an overall phase factor,
the solution is then
\beq
 \Psi (t,r)=\exp\left[\,-i\int_{r_{0}}^{r}{\mathrm d}r' \: a_{0}(t,r')/2 \,\right]
\Psi_{\mathrm  M}(t,r),\label{Psi0}
\eeq
provided $\Psi_{\mathrm  M}(t,r)$ satisfies the nontrivial matrix equation
\beq
\partial_r\Psi_{\mathrm M} =\left[\,
 -i\,\gamma_{5} \,a_{1}/2
 - \left(\gamma^{1}\,\mbox{Re}\chi+ i\gamma^{0}\,\mbox{Im}\chi\right)/r\, \right]
 \Psi_{\mathrm  M},
\label{eq-Psi0}
\eeq
with boundary condition
\beq
\Psi_{\mathrm  M} (t,0)=0.\label{bc-Psi0}
\eeq
The existence of a fermion zero mode for the linear differential equation
(\ref{FZMeq-ND}) is thus equivalent to having a normalizable solution of Eq.
(\ref{eq-Psi0}) with boundary condition (\ref{bc-Psi0}).

Next, we apply a unitary transformation to Eq. (\ref{eq-Psi0}),
\beq
\Psi_{\mathrm  M}\rightarrow\Psi_{\Lambda}\equiv \Lambda^{\dagger}\,
\Psi_{\mathrm  M},\label{unitaryT}\\
\eeq
with the transformation matrix
\beq
\Lambda=-i\,\gamma^{0}\exp [i\,\varphi\, \gamma_{5}/2],\label{Lambda}
\eeq
that diagonalizes the Hermitian matrix ${\cal A}_{H}$ via
${\cal A}_{H}\rightarrow \Lambda^{\dagger} {\cal A}_{H}\Lambda$.
We obtain the following zero-eigenvalue equation for $\Psi_{\Lambda}$:
\begin{mathletters} \label{FZMeq-D}
\begin{eqnarray}
\partial_r \Psi_{\Lambda}&=&({\cal A}_{0}+{\cal A}_{1})\,\Psi_{\Lambda},
\label{FZMeq-D-eq}\\
{\cal A}_{0}&\equiv&\Lambda^{\dagger}\,{\cal A}_{H}\,\Lambda
=\lambda\,\gamma^{1} = - \lambda\,\sigma^3,\label{FZMeq-D-A0}\\
{\cal A}_{1}&\equiv&\Lambda ^{\dagger}\,
                    (-\partial_{r}-i\gamma_{5} \,a_{1}/2)\,\Lambda
={\cal R}\,i\gamma_{5} = {\cal R}\,i\sigma^2  ,\label{FZMeq-D-A1}
\end{eqnarray}
\end{mathletters}
with the further definitions
\beq
\lambda\equiv\rho/r \geq 0, \quad {\cal R}\equiv (a_{1}-\partial_{r}\varphi )/2\mbox{.}
\label{def-lambda-R}
\eeq
It follows immediately from the definition (\ref{chi-polar})
that the matrices ${\cal A}_{0}$ and ${\cal A}_{1}$ are invariant
under the $U(1)$ gauge transformations (\ref{gaugeT-2d}).
Moreover, ${\cal A}_{0}$ and ${\cal A}_{1}$ are real matrices and the solution
$\Psi_{\Lambda}$ can be taken real, up to an overall complex phase
factor. In the following, we take $\Psi_{\Lambda}$ to be strictly
real and drop the subscript $\Lambda$.

For finite-energy background gauge fields with $r$-parity expansions
as given by Eq. (\ref{regular}), one can show that the following limits hold:
\beq
\lim_{r\rightarrow 0} {\cal R}/\lambda=
\lim_{r\rightarrow \infty} {\cal R}/\lambda=0\mbox{.}
\label{local}
\eeq
This demonstrates that the diagonal matrix ${\cal A}_{0}$ determines the local
structure of the solution of the \diffeq~(\ref{FZMeq-D}) in the regions of small and
large $r$ (see also Ref. \cite{V96}).

\subsection{Spinor twist number and twist factor}

Since the solution $\Psi$ of the transformed \zeieq~(\ref{FZMeq-D}) is taken to be real,
one can write $\Psi$ in polar notation,
\beq
\Psi(t,r)\equiv |\Psi(t,r)|\;\exp[\,i\,\gamma_{5}\,\Theta (t,r)]
\left(\begin{array}{c} 0 \\ 1 \end{array}\right),
\label{Psi-polar}
\eeq
where $\Theta\in \mathbb{R}$ measures the relative rotation of the spinor away from the
$\Psi_{2}$-axis in the configuration space of $\Psi$. Recall that
$\gamma_{5}=\sigma^{2}$, so that the exponential factor in Eq. (\ref{Psi-polar})
reads $\openone \cos\Theta + i\sigma^2 \sin\Theta$.

 From Eqs. (\ref{FZMeq-D}) and (\ref{Psi-polar}), one
finds that $\Theta$ and $|\Psi |$  at fixed
time $t$ satisfy the following coupled \diffeqs:
\begin{mathletters}\label{FZMeq}
\begin{eqnarray}
\partial_{r}\Theta &=& {\cal D}[\Theta] +{\cal R},\label{FZMeq-Theta}\\
\partial_{r}|\Psi |&=&\lambda\, |\Psi |\,\cos 2\Theta \mbox{,}\label{FZMeq-Psi}
\end{eqnarray}
\end{mathletters}
with the definitions
\beqa  \label{DRdef}
{\cal D}[\Theta] &\equiv& -\lambda\,\sin 2\Theta  , \quad
{\cal R} \equiv (a_{1}-\partial_{r}\varphi )/2 ,\nonumber\\
\lambda &\equiv& \rho/r \geq 0 \;.
\eeqa
In order to obtain regular behavior at $r=0$,
the solutions of the \diffeqs~(\ref{FZMeq})
must satisfy the following boundary conditions:
\begin{mathletters}
\label{FZMeq-bcs}
\beqa
\Theta (t,0)&=&0 \bmod \pi, \label{BCTheta}\\
|\Psi (t,0)|&=&0  .\label{BCPsi}
\eeqa
\end{mathletters}
More specifically, these boundary conditions are needed because $\lambda$ is
singular at the spatial origin $r=0$; see Eq. (\ref{r-zero}).

The \emph{nonlinearity} of the \diffeq~(\ref{FZMeq-Theta}) originates from the
fact that the linear \diffeq~(\ref{FZMeq-D}) mixes the components of the
spinor $\Psi$. Furthermore, the \diffeq~(\ref{FZMeq-Theta})
involves \emph{only} $\Theta$, whereas Eq. (\ref{FZMeq-Psi}) contains both $\Theta$ and
$|\Psi|$. These two properties of  Eq. (\ref{FZMeq-Theta})
will turn out to be crucial for the results of the present paper.

Remarkably, the nonlinear \diffeq~(\ref{FZMeq-Theta}) for a given time slice $t$
can be transformed into a generalized Riccati equation
\cite{I56,H69} by setting $Y(t,r)$ $=$ $\tan \Theta(t,r)$,
\beq
\partial_{r}  Y - {\cal R}\left(\,1+Y^2\,\right)+2\,\lambda \:Y=0\; .\label{Riccati}
\eeq
The analysis is, however, best carried out with the
nonlinear  \diffeq~in the form as given by Eq. (\ref{FZMeq-Theta}),
where the term ${\cal D}$ is called the ``deviator'' and the term ${\cal R}$ the
``rotator,'' for reasons that will become clear shortly.
Henceforth, we refer to the single \diffeq~(\ref{FZMeq-Theta}), with the
implicit boundary condition  (\ref{BCTheta}), as the ``\tReq.''

Let us consider the asymptotic behavior of the solution $\Theta(t,r)$ of the
\tReq~(\ref{FZMeq-Theta}) at a fixed time $t$.
The deviator ${\cal D}$ dominates, in general,
the right-hand side of Eq. (\ref{FZMeq-Theta}) for large $r$, according to
Eq. (\ref{local}).  For large $r$,
Eq. (\ref{FZMeq-Theta}) can  therefore be approximated by
\beq
\partial_{r}\Theta =-\lambda\,\sin 2\Theta\mbox{.}
\label{FZMeq-asym}
\eeq
The \diffeq~(\ref{FZMeq-asym}) has three types of solutions at a fixed time
slice $t$,
\begin{mathletters}
\begin{eqnarray}
\Theta (t,r)&=&N\,\pi  ,\label{FZMsol-zero}\\[0.2cm]
\Theta (t,r)&=&(N^\prime+1/2)\,\pi,  \label{FZMsol-half}\\[0.2cm]
\tan[\Theta (t,r)]&=&\tan[\Theta(t,r_{0})]\nonumber\\
& & \times \exp\left[-2\int_{r_{0}}^{r}{\mathrm d}r' \: \lambda(t,r')\right]
   \mbox{,}
\label{FZMsol-asym}
\end{eqnarray}
\end{mathletters}
for arbitrary integers $N$ and $N^\prime$.

The nontrivial solution $\Theta(t,r)$ given by Eq. (\ref{FZMsol-asym}) is attracted
toward the value $N\,\pi$ as $r\rightarrow\infty$, since
$\lambda(t,r)$ is non-negative and has a divergent integral toward infinity.
This shows that the ``point'' $\Theta(t,r) =N\,\pi$, with $N\in\mathbb{Z}$,
is \emph{asymptotically stable} in the
solution space of the \diffeq~(\ref{FZMeq-asym}); cf. Ref. \cite{V96}.
For the trivial solution $\Theta(t,r) =(N^\prime+1/2)\,\pi$,
an arbitrarily small deviation will
lead to a nontrivial solution given by Eq. (\ref{FZMsol-asym}), which
asymptotically approaches the value $N''\,\pi$, with $N''\in\mathbb{Z}$.
The ``point'' $\Theta(t,r) =(N^\prime+1/2)\,\pi$, with $N'\in\mathbb{Z}$, is thus
\emph{asymptotically unstable} in the solution space of the
\diffeq~(\ref{FZMeq-asym}).

The solutions of the complete \diffeq~(\ref{FZMeq-Theta})
with boundary condition (\ref{BCTheta})
can thus be classified according to their asymptotic behavior.
At a fixed time slice $t$, there are two classes:
\begin{mathletters}
\begin{eqnarray}
S_{N}(t)&\equiv& \{\Theta (t,r)|\,\lim_{r\rightarrow\infty}
              \Theta(t,r) =N\,\pi \},\label{SN}\\
U_{N^\prime}(t)&\equiv& \{\Theta (t,r)|\,\lim_{r\rightarrow\infty}
               \Theta(t,r) =(N^\prime+1/2)\,\pi \},\label{UN}
\end{eqnarray}
\end{mathletters}
with $N, N^\prime \in\mathbb{Z}$. If $\Theta \in S_{N}$ ($\Theta\in U_{N}$), then
$\Theta$ is asymptotically stable (unstable) in the solution space of
the \diffeq~(\ref{FZMeq-Theta}).

For any solution $\Theta(t,r) \in U_{N}(t)$ with arbitrary integer $N$,
there necessarily exists a fermion zero mode,
as follows from Eqs. (\ref{FZMeq-Psi}) and (\ref{BCPsi}).
[It is clear that the normalizability condition (\ref{norm}) of the fermion
zero mode requires the asymptotics of Eq. (\ref{UN}).]
Therefore, it \emph{suffices} to study the \tReq~(\ref{FZMeq-Theta}) in order to
determine the existence of a fermion zero mode at a particular time $t$.

At this moment, we can explain the use of the terms ``deviator'' and
``rotator'' in the \tReq~(\ref{FZMeq-Theta}). The observation from Eqs.
(\ref{FZMeq-asym}) and (\ref{FZMsol-asym}) is
that ${\cal D}$ pulls $\Theta(t,r)$ toward
the value $N\,\pi$ as $r\rightarrow\infty$. In other words, it leads to a
deviation of $\Theta(t,r)$ from
the special path approaching the value $(N^\prime+1/2)\,\pi$ as $r\rightarrow\infty$,
for which a fermion zero mode exists.
This is the reason for calling the term ${\cal D}$ in Eq.
(\ref{FZMeq-Theta}) the ``deviator.''
In the absence of the deviator ${\cal D}$ over the interval
$[r_0,r_1]$, say, one observes from Eq. (\ref{FZMeq-Theta}) that
${\cal R}$ generates
a simple rotation of the spinor by the angle $\Delta\Theta
=\int_{r_{0}}^{r_1} {\mathrm d}r' \: {\cal R}(r')$.
This is then the reason for calling the term ${\cal R}$ in Eq.
(\ref{FZMeq-Theta})  the ``rotator.''
The  fermion zero mode solutions will be discussed further in the next subsection.
Here, we continue the discussion of the \tReq~from a more general viewpoint.

Using analyticity and the Cauchy-Lipschitz existence and uniqueness
theorem for ordinary differential equations \cite{I56,H69,V96}, it can
be shown that the
solution $\Theta(t,r)$ of the \tReq~(\ref{FZMeq-Theta}) with boundary condition
$\Theta (t,0)=0$ is unique. For the regular finite-energy gauge fields
considered, it can also be shown that the solution $\Theta(t,r)$ is bounded.

The uniqueness  of the solution $\Theta(t,r)$  and its
asymptotic behavior allow us to \emph{classify} the gauge
field background at one particular time $t$ by the quantity
\beq
N_{\Theta}(t)\equiv [\,\Theta (t,\infty )-\Theta(t,0)\,]/\pi,\label{twistN}
\eeq
which can take integer or half-odd-integer values. According to the definition
(\ref{Psi-polar}),
the mapping ${\cal G}(t,r)$ $\equiv$ $\exp[i\,\gamma_{5}\,\Theta (t,r)]$
$\in$ $SO(2)$ gives the twisting of the spinor in the configuration space of $\Psi$
for fixed time $t$. We therefore call $N_{\Theta}(t)$ the ``spinor twist number.''

It is now convenient to characterize a time-dependent  spherically symmetric $SU(2)$
gauge field background by the change of spinor twist number between initial and
final configurations. Henceforth, we call
\beq
\Delta N_{\Theta}[\,t_f,t_i\,]\equiv N_{\Theta}(t_{f})-N_{\Theta}(t_{i})\label{twistF}
\eeq
the ``twist factor'' of the  spherically symmetric  $SU(2)$ gauge field
over the time interval $[\,t_i,t_f\,]$, with $t_i < t_f$.

It is important to realize that the twist factor $\Delta N_{\Theta}$ measures an
\emph{intrinsic property} of the $SU(2)$ gauge field configuration. Formally,
Eqs. (\ref{FZMeq-D}a), (\ref{Psi-polar}), (\ref{twistN}), and
(\ref{twistF}) give
\begin{mathletters}   \label{twistfactorformal}
\begin{eqnarray}
\Delta N_{\Theta}[\,t_f,t_i\,]&=&\frac{1}{\pi}
\int_{0}^{\infty}{\mathrm d}r\int_{t_{i}}^{t_{f}}{\mathrm d}t\;
\frac{\partial}{\partial t}\left(\frac{\partial}{\partial r}
\Theta (t,r)\right),\\[0.2cm]
\Theta (t,r)&\equiv&
-\frac{1}{2}\,\mbox{Tr}\left(\,i\sigma^{2}\ln\left\{
\lim_{\epsilon\rightarrow 0}\, \epsilon \, {\cal P}\exp
\left[\,\int_{\epsilon}^{r}{\mathrm d}r'\;
\{ {\cal A}_{0}(t,r')+{\cal A}_{1}(t,r')\}\right]\right\}\right),
\end{eqnarray}
\end{mathletters}
\noindent where ${\cal P}$ represents path ordering. Here, ${\cal A}_{0}(t,r)$ and
${\cal A}_{1}(t,r)$ are defined by Eqs. (\ref{FZMeq-D}b,c), in terms of the
(1+1)-dimensional gauge field functions $\rho(t,r)$, $\varphi(t,r)$,
and $a_1(t,r)$. Whether or not there exists a more direct way to
obtain $\Delta N_{\Theta}$ remains an open question.

\subsection{Necessary condition for fermion zero modes}

With the results of the previous subsection, it is possible
to find a necessary condition for the existence of fermion zero modes
at a particular time $t$. We first introduce the following
diagnostic:
\beqa  \label{Kpm}
{\cal K}_{\pm}(t) &\equiv&\int_{D_{\pm}(t)}{\mathrm d}r \: {\cal R}(t,r)
\nonumber\\[0.2cm]
                  &\equiv& \int_{0}^{\infty} {\mathrm d}r \:
                  \theta[\,\pm\,{\cal R}(t,r)\,]\: {\cal R}(t,r)\;,
\eeqa
with the domains of positive or negative values of ${\cal R}(t,r)$ defined by
\beq
D_{\pm}(t)\equiv\{r|\: \mbox{sgn}[{\cal R}(t,r)]=\pm 1\}\subseteq [0,\infty )
\eeq
and $\theta$ the usual step function, $\theta[x]$ $=$ $0$ for $x$ $<$ $0$ and
$\theta[x]$ $=$ $1$ for $x$ $>$ $0$.
Note that, by definition, ${\cal K}_{+}$ $\geq$ $0$ and ${\cal K}_{-}$
$\leq$ $0$. Note also that the rotator ${\cal R}(t,r)$ from Eq. (\ref{DRdef})
is entirely defined in terms of the background fields
$a_1(t,r)$ and $\varphi(t,r)$ $\equiv$ $\arg\,\chi(t,r)$.

Consider the \tReq~(\ref{FZMeq-Theta}) with boundary condition $\Theta(t,0)=0$.
The integration of ${\cal R}(t,r)$ over the domain $D_{+}$($D_{-}$) then
accounts for the rotation of the spinor in the ``$+$"(``$-$") direction.
But the  deviator  ${\cal D}$, for values $\Theta \in (-\pi/2,\pi /2)$,
brakes the rotation forced by the rotator  ${\cal R}$.
The crucial point, now, is that in order to have a fermion zero mode at time
$t$ the total action of the rotator should overcome the resistance from the
deviator in the region $-\pi /2 <\Theta <\pi /2$, so that the solution
$\Theta(t,r)$ ends up with $|\,\Theta |\geq \pi /2$ at $r=\infty$.
A \emph{necessary} condition for the existence of a fermion zero mode at
one particular time $t$ is, therefore,
\beq
{\cal K}_{\mathrm max}(t) \equiv
         \max\left[\,{\cal K}_{+}(t)\,,|\,{\cal K}_{-}(t)\,|\,\right]\geq \pi /2 \mbox{.}
\label{Kmax}
\eeq

Having established the necessary condition (\ref{Kmax}), it would
certainly be interesting to obtain also a necessary and sufficient condition
for the existence of a fermion zero mode in a given static gauge field background.
But, without further input, it appears difficult to find such a condition.
For this reason, we turn in the next section to the role of time-dependent,
continuous gauge field backgrounds.

\section{Spectral flow}

In this section, we consider the spectrum of the effective Dirac Hamiltonian
(\ref{Hamiltonian}) for  time-dependent spherically symmetric $SU(2)$
gauge fields. In Sec. IV A, we derive a relation, Eq. (\ref{spectral-local}),
between level crossing and the change of winding number or spinor
twist number over an infinitesimal time interval.  From this
result, we obtain in Sec. IV B the appropriate  relation, Eq. (\ref{SelectionRule}),
for the spectral flow over a finite time interval.
Section IV A is rather technical and may be skipped on a first reading.

\subsection{Level crossing from changes in winding and twist numbers}

\subsubsection{Perturbative expansion}

We start from the \tReq~(\ref{FZMeq-Theta}),
with boundary condition (\ref{BCTheta}),
at a particular time $t=t^{*}$ and study
the change of the solution $\Theta (t,r)$ in the
neighborhood of $t=t^{*}$. For finite $r$
and $t=t^{*}\pm\epsilon$ with $\epsilon$ an arbitrarily small positive constant,
one can expand the background fields as follows
\begin{mathletters} \label{background-t}
\begin{eqnarray}
\lambda (t^{*}\pm\epsilon ,r)&=&\lambda (t^{*},r)\pm\epsilon\,
\left.\partial_{t}\lambda \right|_{t=t_{\pm}^{*}}+
{\mathrm O}(\epsilon^2), \\
{\cal R}(t^{*}\pm\epsilon ,r)&=&{\cal R}(t^{*},r)\pm\epsilon\,
\left.\partial_{t}{\cal R}\right|_{t=t_{\pm}^{*}}+
{\mathrm O}(\epsilon^2),
\end{eqnarray}
\end{mathletters}
where the upper and lower time derivatives of the background fields are defined
by
\beqa \label{updownderivatives}
\left.\partial_{t}\lambda (t,r)\right|_{t=t_{+}^{*}}&\equiv&
\lim_{t\,\downarrow\, t^{*}}\,\partial_{t}\lambda (t,r),  \nonumber\\
\left.\partial_{t}\lambda (t,r)\right|_{t=t_{-}^{*}}&\equiv&
\lim_{t\,\uparrow\, t^{*}}\,\partial_{t}\lambda (t,r),
\eeqa
and similarly for $\partial_{t}{\cal R}$.
The solution $\Theta(t,r)$ for $t=t^{*}\pm\epsilon$ can be written as
\beq
\Theta (t^{*}\pm\epsilon ,r)=\Theta_{\pm}(t^{*},r)\pm\epsilon\,
f_{1}(t^*,r)+{\mathrm O}(\epsilon^2)\mbox{.}\label{Theta-t}
\eeq
The function $f_1(t,r)$ is continuous
at $t=t^{*}$, but the first term on the \rhs~of Eq. (\ref{Theta-t})
allows for a discontinuity. For the moment, we consider the functions
$\Theta_{\pm}(t^{*},r)$ in  Eq. (\ref{Theta-t}) to be equal.

By substituting Eqs. (\ref{background-t}) and (\ref{Theta-t}) into the
\tReq~(\ref{FZMeq-Theta}), one obtains to first order in $\epsilon$
the following linear \diffeq~for $f_{1}$:
\beq
\left[ \,\partial_{r}+2\,\lambda (t^*,r)\cos
2\Theta (t^{*},r)\,\right]\,f_{1}(t^*,r)= j_{1}(t^{*},r)\label{eq-f1},
\eeq
with the definition
\beq
j_{1}(t^{*},r)\equiv \left.\partial_{t}{\cal
R}\right|_{t=t^{*}}-\left.\partial_{t}\lambda
\right|_{t=t^{*}}\sin 2\Theta ,\label{eq-j1}
\eeq
and boundary condition
\beq
f_{1}(t^*,0)=0\mbox{.}
\eeq

The solution of the \diffeq~(\ref{eq-f1}) is found to be
\beq
f_{1}(t^*,r)= {\cal J}_{1}(t^*,r)/\, |\Psi (t^{*},r)|^2, \label{sol-f1}
\eeq
with the definition
\beq
{\cal J}_{1}(t^*,r)\equiv \int_{0}^{r}{\mathrm d}r' \:
|\Psi (t^{*},r' )|^2 \,j_{1}(t^{*},r')\:.\label{def-J1}
\eeq
Here, we have used the solution of Eq. (\ref{FZMeq-Psi}),
\beq
|\Psi (t,r)| \propto \exp\left[\,\int_{r_{0}}^{r}\rmd r' \:
\lambda (t,r')\cos 2\Theta (t,r')\right],
\label{sol-Psi}
\eeq
to obtain Eq. (\ref{sol-f1}) in the form shown.

The function $j_{1}(t,r)$ is continuous at $t=t^{*}$ for smooth
background  fields $\lambda(t,r)$ and ${\cal R}(t,r)$, which implies that the
solution $f_{1}(t,r)$ of Eq. (\ref{eq-f1}) is also continuous at $t=t^{*}$.
If, on the other hand, the time slice $t=t^{*}$
corresponds to a local change of the gauge field winding number $N_\chi$, the
partial derivatives of $\lambda(t,r)$ and ${\cal R}(t,r)$ are not
well-defined at $t=t^{*}$, as will be shown later. In this case, the
function $j_{1}(t,r)$ is not well-defined either, which affects
the continuity of the solution $f_{1}(t,r)$ of Eq. (\ref{eq-f1}).
In Sec. IV A 3, we will show that the function $f_1(t,r)$ can be taken to be
continuous at $t=t^{*}$, provided possible discontinuities
are accounted for by the leading terms
$\Theta_{-}(t^{*},r)$ and $\Theta_{+}(t^{*},r)$ in Eq. (\ref{Theta-t}).

\subsubsection{Time-differentiable $\lambda$ and ${\cal R}$}

Consider a particular time slice $t=t^{*}$, for which the $\lambda(t,r)$
and ${\cal R}(t,r)$ fields are differentiable with respect to time,
\begin{mathletters}  \label{cont-t}
\beqa
\left.\partial_{t}\lambda
(t,r)\right|_{t=t^{*}_{+}}&=&\left.\partial_{t}\lambda
(t,r)\right|_{t=t^{*}_{-}}, \\
\left.\partial_{t}{\cal R}
(t,r)\right|_{t=t^{*}_{+}}&=&\left.\partial_{t}{\cal R}
(t,r)\right|_{t=t^{*}_{-}}\mbox{,}
\eeqa
\end{mathletters}
with upper and lower time derivatives as defined in  Eq. (\ref{updownderivatives}).

First, suppose that there is no fermion zero mode at $t=t^{*}$,
so that $\Theta(t^{*},r)$ $\rightarrow$ $N\,\pi$ as $r$ $\rightarrow$ $\infty$.
For large $r$ and using  Eq. (\ref{local}),
one then obtains from Eq. (\ref{eq-f1}) the results
$\lim_{r\rightarrow\infty}j_{1}/\lambda =0$  and $f_{1} \propto r^{-2}$,
which imply
\beq
\lim_{r\rightarrow\infty}f_{1}(t^{*},r)=0.\label{f1-asym}
\eeq
This indicates that there is no change of the
asymptotic behavior of $\Theta(t,r)$ in the neighborhood of $t=t^{*}$,
which corresponds to having a constant spinor twist number (\ref{twistN}) at $t=t^*$,
\beq
\delta N_{\Theta}|_{t=t^{*}}\equiv N_{\Theta}(t^{*}+\epsilon )-
N_{\Theta}(t^{*}-\epsilon )= 0 \label{twist-zero} \mbox{.}
\eeq
[We reserve the notation $\Delta N_{\Theta}$ for the \emph{global} change of
spinor twist number; see Eq. (\ref{twistF}). Of course, $\delta N_{\Theta}$ is in no
way ``infinitesimal,'' see Eqs. (\ref{twist-plusmin}) below.]

Next, consider the case of having a normalized fermion zero mode at $t=t^{*}$
with $\Theta (t^{*},r)\in U_{N}$, that is, belonging to the ``unstable''
class of solutions (\ref{UN}). Since the solution $\Theta (t^{*},r)\in U_{N}$ is
asymptotically unstable, one observes from Eq. (\ref{FZMsol-asym}) that a small
positive [negative] perturbation of $\Theta$ at large $r$ leads to
$\Theta (t^{*},r)\in S_{N+1}$ [$\Theta (t^{*},r)\in S_{N}$].  From Eqs.
(\ref{Theta-t}), (\ref{sol-f1}), and (\ref{def-J1}) one deduces that the
fermion zero mode at $t=t^{*}$ sits at a \emph{bifurcation point} for different
$N_{\Theta}$'s. In fact, the local change of spinor twist number
\beq \label{deltaNTheta}
\delta N_{\Theta}|_{t=t^{*}} \equiv N_{\Theta}(t^{*}+\epsilon )-
                                    N_{\Theta}(t^{*}-\epsilon )
\eeq
is given by
\beq                 \label{twist-plusmin}
\delta N_{\Theta}|_{t^{*}}= \left\{ \begin{array}{ll} 
+1  & \mbox{for} \quad {\cal J}_{1}(t^{*},\infty )>0, \\
-1  & \mbox{for} \quad {\cal J}_{1}(t^{*},\infty )<0,
\end{array}\right.
\eeq
as long as ${\cal J}_{1}(t^{*},\infty ) \neq 0$. [The special case of
${\cal J}_{1}(t^{*},\infty )=0$ will be discussed in Sec. IV A 4.]
For an elementary discussion of bifurcation theory; see Ref.  \cite{V96}.

Having a fermion zero mode at $t=t^{*}$, we are especially
interested in the time-gradient of the fermion energy eigenvalue at
$t=t^{*}$, in order to check for level crossing. The
time-gradient of the energy eigenvalue of the Dirac Hamiltonian at
$t=t^{*}$ is calculated up to the first order in $\epsilon$:
\beqa
\left.\dEdt \right|_{t=t^{*}}&=& \:
\left<\Psi_{l}(t^{*},r) \left|\frac{\partial
H}{\partial t}(t^{*})\right|\Psi_{l}(t^{*},r)\right> \nonumber\\[0.2cm]
& =& {\cal J}_{1}(t^{*},\infty ), \label{crossing-J1}
\eeqa
where ${\cal J}_{1}$ is defined by Eq. (\ref{def-J1}) and $\Psi_{l}(t^{*},r)$
represents the (nondegenerate) normalized fermion zero mode at $t=t^{*}$ in the
two-component  spinor notation of Eq. (\ref{notation-2d}).
The expectation value used in Eq. (\ref{crossing-J1}) is defined by
\beqa
& &\left<\Psi_{l}(t^{*},r) \left|\,O(t^*)\, \right|\Psi_{l}(t^{*},r)\right>\nonumber\\
& &\equiv \int_{0}^{\infty} \rmd r\;\Psi_{l}(t^{*},r)^\dagger
                                     \,O(t^*)\, \Psi_{l}(t^{*},r)\;,
\eeqa
for an arbitrary  time-dependent Hermitian operator $O(t)$.

 From Eqs. (\ref{twist-plusmin})  and (\ref{crossing-J1}), we obtain
\beq
{\mathrm sgn}\left[\left.\dEdt
\right|_{t=t^{*}}\right]=
  \left.\delta N_{\Theta}\right|_{t=t^{*}} \in \{-1,+1 \}\mbox{,}
\label{crossing-twist}
\eeq
with the implicit limit $\epsilon \rightarrow 0$ on the right-hand side.
This establishes the relation between level crossing and the change of spinor
twist number, for the case that  $\lambda(t,r)$ and ${\cal R}(t,r)$ are
time-differentiable at $t=t^{*}$ and generic for times close to it
[so that ${\cal J}_{1}(t^{*},\infty ) \neq 0$].

\subsubsection{Time-nondifferentiable $\lambda$ and ${\cal R}$}

Now, consider a gauge field background for which $\chi(t,r)$ vanishes at the
spacetime point $(t^{*},r^{*})$ and the winding number $N_{\chi}$ as
defined
in Eq. (\ref{windingN}) changes from a value $N$ to $N+\delta
N_{\chi}|_{t=t^{*}}$.
Generally, the fields $\lambda(t,r)$ and ${\cal R}(t,r)$ are
not differentiable with respect to time:
\begin{mathletters}  \label{discont-t}
\beqa
\left.\partial_{t}\lambda
(t,r)\right|_{t=t^{*}_{+}}&\neq&\left.\partial_{t}\lambda
(t,r)\right|_{t=t^{*}_{-}}, \\
\left.\partial_{t}{\cal R}
(t,r)\right|_{t=t^{*}_{+}}&\neq&\left.\partial_{t}{\cal R}
(t,r)\right|_{t=t^{*}_{-}}\mbox{,}
\eeqa
\end{mathletters}
with upper and lower time derivatives as defined in  Eq. (\ref{updownderivatives}).

Let us have a closer look at the discontinuities of the
time derivatives of $\lambda(t,r)$ and ${\cal R}(t,r)$ at $t=t^{*}$.
First, one observes from the definition (\ref{DRdef}) of the $\lambda(t,r)$ field
that the  time derivative of $\lambda(t,r)$ is not well-defined at the spacetime point
$(t^{*},r^{*})$:
\beqa
\left.\partial_{t}\lambda (t,r^{*})\right|_{t=t_{+}^{*}}&=&
-\left.\partial_{t}\lambda (t,r^{*})\right|_{t=t_{-}^{*}}\nonumber\\
&=&\left|\partial_t \chi (t^{*},r^{*})\right|/r^{*}.
\label{discont-lambda}
\eeqa

Second, introduce the gauge-invariant function $\mbox{\sf R}(t^*,r)$ defined by
\begin{eqnarray}
\mbox{\sf R}(t^*,r)&\equiv &\lim_{\epsilon\rightarrow 0}\left[{\cal
R}(t^{*}+\epsilon ,r)-{\cal R}(t^{*}-\epsilon ,r)\right] \nonumber\\
&=& \lim_{\epsilon\rightarrow
0}\left[\left.\partial_{r}\varphi\right|_{t=t^{*}-\epsilon}
-\left.\partial_{r}\varphi\right|_{t=t^{*}+\epsilon}\right]/\,2\mbox{.}
\label{def-R}
\end{eqnarray}
Taylor expanding $\partial_{r}\varphi(t,r)$ with respect to both $t$ and $r$
in the vicinity of the spacetime point $(t^{*},r^{*})$ and using
the fact that $\alpha(t,r)\rightarrow 0$ and $\beta(t,r)\rightarrow 0$ as
$(t,r)\rightarrow (t^{*},r^{*})$, one finds that $\mbox{\sf R}(t^*,r)$
shoots up to infinity at $r=r^{*}$, whereas it drops to zero for
$r\neq r^{*}$. Taking the gauge condition
$\chi (t,0)=\chi (t,\infty )=1$, one readily proves that
\begin{eqnarray}
& &\int_{0}^{\infty}{\mathrm d}r\;\mbox{\sf R}(t^*,r)
=-\pi\left.\lim_{\epsilon\rightarrow 0} \delta N_{\chi}\right|_{t=t^{*}} ,
\end{eqnarray}
with the local change of winding number defined as
\begin{eqnarray} \label{deltaNchi}
& &\left.\delta N_{\chi}\right|_{t=t^{*}}\equiv N_{\chi}(t^{*}+\epsilon
)-N_{\chi}(t^{*}-\epsilon )\mbox{.}
\end{eqnarray}
This shows that the function $\mbox{\sf R}(t^*,r)$ is proportional to a
Dirac delta-function centered at $r=r^{*}$,
\beq
\mbox{\sf R}(t^*,r)=-\pi\,\lim_{\epsilon\rightarrow 0}\delta N_{\chi}|_{t=t^{*}}\;
\delta (r-r^{*})\mbox{.}
\label{R-dirac}
\eeq
The nonvanishing \rhs~of Eq. (\ref{R-dirac}) for $r=r^{*}$
implies that the  time derivative of the rotator ${\cal R}(t,r)$ is
not well-defined at the spacetime point $(t^{*},r^{*})$.

The delta-function-like behavior (\ref{R-dirac}) of $\mbox{\sf R}$
can be used to derive the effect of the change of the gauge field winding
number on the change of the spinor twist number at $t=t^{*}$.
Start by defining
\beq
\Theta_{-}(t^{*},r) \equiv \lim_{t\,\uparrow\, t^{*}}\,\Theta(t,r).
\label{Theta-minus}
\eeq
Then, if $\Theta_{-}(t^{*},r)$ belongs to the class $S_{N}$, for some integer
$N$, one deduces from Eqs. (\ref{FZMeq-Theta}),
(\ref{def-R}), and (\ref{R-dirac}) that
\beqa
\Theta_{+}(t^{*},r)&\equiv&\lim_{t\,\,\downarrow\,\, t^{*}}\,\Theta(t,r)\nonumber\\
&=&\Theta_{-}(t^{*},r) -\pi\,\lim_{\epsilon\rightarrow 0}\delta N_{\chi}|_{t=t^{*}}\,
\theta (r-r^{*}),
\label{Theta-step}
\eeqa
where $\theta$ is the usual step function. This
implies that the solution changes from one class to another, as $t$
crosses the value
$t^*$. Specifically, if the earlier solution
$\Theta_{-}(t^{*},r)$ belongs to $S_{N}$, then the later solution
$\Theta_{+}(t^{*},r)$ belongs to $S_{N-\delta N_{\chi}}$.

Equation (\ref{Theta-step})
shows that the change of the gauge field winding number $\delta
N_{\chi}|_{t=t^{*}}$ causes the change of the spinor twist number at
$t=t^{*}$ to be given by
\beq
\delta N_{\Theta ; {\mathrm A}}|_{t=t^{*}}=-\delta N_{\chi}|_{t=t^{*}},
\label{dN-Psi-A}
\eeq
regardless of the existence of a fermion zero mode at $t=t^{*}$
(this contribution is labeled A).

Before we continue with the evaluation of  the change of spinor twist number,
we need to address the continuity issue for the
solution $f_1$ of Eq. (\ref{eq-f1}).
According to Eqs. (\ref{discont-lambda}) and
(\ref{R-dirac}),  the  time derivatives of the gauge-invariant $\lambda(t,r)$
and ${\cal R}(t,r)$ fields are not well-defined at the spacetime point
$(t^{*},r^{*})$. This implies that the function $j_{1}(t,r)$, as defined in
Eq. (\ref{eq-j1}), does not have a well-defined value either. Note that all these
problems can be traced to the discontinuity of the unitary matrix $\Lambda$ in Eq.
(\ref{Lambda}), which, in turn, is caused by the ill-defined argument
$\varphi(t,r)$ of the field $\chi(t,r)$ at the spacetime point $(t^{*},r^{*})$ where
$\chi(t,r)$ vanishes.
By performing the inverse unitary transformation
of Eq. (\ref{unitaryT}), $\Psi\rightarrow \Psi_{l}\equiv \Lambda\Psi$, one
finds that  the $j_1$ ``expectation value'' (\ref{def-J1}) is again given by
\beq
{\cal J}_{1}(t^{*},R)=\left<\Psi_{l} (t^{*},r)
\left|\frac{\partial H}{\partial
t}(t^{*})\right|\Psi_{l} (t^{*},r)\right>_R,
\label{cont-J1}
\eeq
with the implicit integral over $r$ on the \rhs~running over $[0,R\,]$.
The \rhs~of Eq. (\ref{cont-J1})
has a well-defined value at $t=t^{*}$ for smooth background fields
$a_{0}(t,r)$, $a_{1}(t,r)$, and $\chi(t,r)$. This shows that the discontinuity of
the function $j_{1}(t,r)$ at $(t^{*},r^{*})$, caused by an ill-defined function
$\varphi(t,r)$,
can be absorbed into the local change of the spinor twist number $\delta
N_{\Theta ; {\mathrm A}}|_{t=t^{*}}$ via the relation (\ref{dN-Psi-A}).

With the discontinuity of $j_{1}(t,r)$ at $(t^{*},r^{*})$ absorbed into
the local change of the spinor twist number $\delta N_{\Theta ; {\mathrm
A}}|_{t=t^{*}}$, the solution $f_{1}(t,r)$ of Eq. (\ref{eq-f1}), explicitly
given by Eq. (\ref{sol-f1}), takes a well-defined value at $t=t^{*}$ and may
produce a local change of spinor twist number that is not associated with
the local change of gauge field winding number.

Finally, we are  ready to consider the additional effect on
the local change of the spinor twist number
due to the presence of a fermion zero mode  at $t=t^{*}$
(this contribution will be labeled B). Since ${\cal J}_{1}(t^{*},\infty)$ as given
by Eq. (\ref{cont-J1}) has a well-defined value at $t=t^{*}$, we have a
unique time gradient for the level crossing,
\beq\label{crossing-cont}
\left.\dEdt \right|_{t=t^{*}+\epsilon}=
\left.\dEdt \right|_{t=t^{*}-\epsilon}={\cal J}_{1}(t^{*},\infty)\mbox{.}
\eeq
Now, generic background gauge fields with  ${\cal J}_{1}(t^{*},\infty
)\neq 0$ produce the following change of spinor twist number at $t=t^{*}$
[see Eqs. (\ref{twist-plusmin}) above]:
\beq
\delta N_{\Theta ; {\mathrm B}}|_{t=t^{*}}={\mathrm sgn}
\left[{\cal J}_{1}(t^{*},\infty )\right],
\label{dN-Psi-B}
\eeq
in addition
to the contribution $\delta N_{\Theta; {\mathrm A}}|_{t=t^{*}}$ given
by Eq. (\ref{dN-Psi-A}).
[Remark that $\delta N_{\Theta; {\mathrm B}}$ $=$ $\pm \,1$
for the generic case considered.]
According to Eq. (\ref{crossing-cont}), the level
crossing at $t=t^{*}$ is determined by ${\mathrm sgn}\left[{\cal
J}_{1}(t^{*},\infty )\right]$. We therefore deduce the following relation
between level crossing and the change of the spinor twist number:
\beqa
{\mathrm sgn}\left[\left.\frac{{\mathrm d}E}{{\mathrm d}t}\right|_{t=t^{*}}\right]
&=&\delta N_{\Theta; {\mathrm B}}|_{t=t^{*}} \nonumber\\
&=&\delta N_{\Theta}|_{t=t^{*}}-\delta N_{\Theta; {\mathrm A}}|_{t=t^{*}},
\label{crossing-dN-AB}
\eeqa
where the total change of the spinor twist number at $t=t^{*}$ is given by
the sum of both contributions
\beq
\delta N_{\Theta}|_{t=t^{*}}=\delta N_{\Theta; {\mathrm A}}|_{t=t^{*}}
+\delta N_{\Theta; {\mathrm B}}|_{t=t^{*}}\mbox{.}
\eeq

Combining Eqs. (\ref{dN-Psi-A}) and (\ref{crossing-dN-AB}), we
find that the local spectral flow
${\cal F}[\,t^{*}+\epsilon, t^{*}-\epsilon \,]$ is given in terms of the local
winding factor and twist factor,
\beq
{\cal F}[\, t^{*}+\epsilon, t^{*}-\epsilon \,]=\delta
N_{\chi}|_{t=t^{*}}+\delta N_{\Theta}|_{t=t^{*}}\mbox{,}
\label{spectral-local}
\eeq
which is the main result of the present subsection. Here,
$\delta N_{\chi}$ and $\delta N_{\Theta}$ are defined by Eqs.
(\ref{deltaNchi}) and (\ref{deltaNTheta}), respectively,
and $\epsilon$ is a positive infinitesimal.

\subsubsection{Special and generic gauge field backgrounds}

Let us, finally, discuss the case of having
a fermion zero mode at $t=t^{*}$, for which
the ``expectation value'' ${\cal J}_{1}(t^{*},\infty )$ vanishes,
\beqa
{\cal J}_{1}(t^{*},\infty )&\equiv& \int_{0}^{\infty}{\mathrm d}r \: |\Psi
(t^{*},r)|^2 \left.\left[\partial_{t}{\cal
R}-\partial_{t}\lambda \,\sin 2\Theta \right]\right|_{t=t^{*}}\nonumber\\
&=&0\mbox{.}
\label{J1-zero}
\eeqa
This implies ${\mathrm d}E/{\mathrm d}t|_{t=t^{*}}=0$, according to Eq.
(\ref{crossing-cont}). A nonvanishing ${\mathrm d}^2 E/{\mathrm
d}t^2 |_{t=t^{*}}$, with vanishing first-order derivative, now corresponds
to the absence of level crossing (the
fermion energy eigenvalue just touches $E=0$ at $t=t^{*}$). But a
nonvanishing ${\mathrm d}^3 E/{\mathrm d}t^3 |_{t=t^{*}}$,
with vanishing  first- and second-order derivatives, again has level crossing.
One therefore needs to find the first nonvanishing derivative of the
energy corresponding to the fermion zero mode, in order to determine whether
or not level crossing occurs.

So far, we have considered only the solution $\Psi (t,r)$ of the
zero-eigenvalue equation (\ref{FZMeq-ND-A}),
not the full spectrum (\ref{Hamiltonianeq}) of
the time-dependent Dirac Hamiltonian.
This makes it impossible to obtain the relation
between the local change of the spinor twist number and the
time derivatives of $E(t)$ at $t=t^{*}$  beyond the leading-order
approximation. Still, the relation (\ref{spectral-local}) can be shown to
hold for generic $SU(2)$ gauge field backgrounds, since background fields with
${\cal J}_{1}(t^{*},\infty )=0$ form a class of measure
zero.

Start from the gauge-invariant background fields $\lambda(t,r)$ and ${\cal
R}(t,r)$ at the time slice $t=t^{*}$ where the fermion zero mode resides, with
profile functions $\Psi (t^{*},r)$ and $\Theta(t^{*},r)$. Under
an infinitesimal time shift $t=t^{*}\rightarrow t^{*}+\epsilon$, generic
$\lambda(t,r)$ and ${\cal R}(t,r)$ vary according to Eq.
(\ref{background-t}), with independent first-order coefficients (possible
differences above and below $t=t^*$ are not important for the present
argument).
The change with time of these background fields is to first order in $\epsilon$
\beq \delta \lambda \equiv \epsilon \;
\partial_{t}\lambda |_{t=t^{*}}, \quad
\delta {\cal R}\equiv
\epsilon \; \partial_{t}{\cal R}|_{t=t^{*}}.\label{delta-lambda-R}
\eeq
Next, define the following ``expectation value'' of $\delta \lambda\,$:
\beqa
<\delta \lambda >_{t=t^{*}}&\equiv& \int_{0}^{\infty}{\mathrm d}r
 \: |\Psi(t^{*},r)|^2\, \left.\,\delta\lambda (t,r)\right|_{t=t^{*}}
 \nonumber\\&& \times \sin 2\Theta(t^{*},r),
\eeqa
with an integration measure weighted by the known function $\sin
2\Theta (t^{*},r)$. The analogous ``expectation value'' of $\delta {\cal R}$ is
\beq <\delta {\cal R}>_{t=t^{*}}\equiv
\int_{0}^{\infty}{\mathrm d}r \: |\Psi (t^{*},r)|^2 \left.\,\delta {\cal
R}(t,r)\right|_{t=t^{*}},
\eeq
but without extra weight function.

Now, recall that the functional ${\cal J}_{1}(t^{*},\infty )$ as given by Eq.
(\ref{J1-zero}) is proportional to the difference of these
``expectation values'' in leading order,
\beqa
\epsilon\, {\cal J}_{1}(t^{*},\infty )&=& \;
<\delta {\cal R}>_{t=t^{*}}-<\delta \lambda >_{t=t^{*}}\nonumber\\
&&+\;{\mathrm O} (\epsilon^2)\,.
\eeqa
Near the origin of the two-dimensional space spanned by
$X$ $\equiv$ $<\delta {\cal R}>_{t=t^{*}}$ and
$Y$ $\equiv$ $<\delta \lambda >_{t=t^{*}}$, the
class of background gauge fields with ${\cal J}_{1}(t^{*},\infty )=0$
therefore coincides with the one-dimensional subspace
\beq
\{(X,Y)|\; X=Y \},
\eeq
which is of measure zero. This shows that  the quantity
${\cal J}_{1}(t^{*},\infty )$ is nonzero for generic background gauge
fields away from $t=t^{*}$ and that the relation
between the local change of the spinor twist number and level crossing
as given by Eqs. (\ref{crossing-twist}) and (\ref{crossing-dN-AB}) holds
in general.

\subsection{Relation between spectral flow and $SU(2)$ gauge field background}

The results of the previous subsection can be summarized as follows.
For \emph{generic} regular bosonic fields of the effective
(1+1)-dimensional theory (\ref{S-2d}), the spectral flow
${\cal F}[\,t_f,t_i\,]$ for the time interval  $[\,t_i,t_f\,]$
is given by the sum of the winding factor (\ref{windingF}) and twist factor
(\ref{twistF}):
\beq
{\cal F}[\,t_f,t_i\,]=\Delta N_{\chi}[\,t_f,t_i\,]+\Delta
N_{\Theta}[\,t_f,t_i\,]\mbox{.}
\label{SelectionRule}
\eeq
This result is simply the grand total of all level crossings (\ref{spectral-local}).

As mentioned before, the spectral flow ${\cal F}$ in Eq. (\ref{SelectionRule})
is defined as the number of eigenvalues of the effective Dirac
Hamiltonian (\ref{Hamiltonian}) that cross zero from below minus the number of
eigenvalues
that cross zero from above, for the time interval $[\,t_i,t_f\,]$
with $t_i < t_f$. The quantity ${\cal F}$ is an integer, by definition. But
the winding factor $\Delta N_{\chi}$
and the twist factor $\Delta N_{\Theta}$ also take integer values in general.

Let us, nevertheless, discuss the special cases for which relation
(\ref{SelectionRule}) is not applicable.
The spectral flow ${\cal F}$ from  Eq. (\ref{SelectionRule}) would not have a
well-defined integer value if the gauge field winding number $N_{\chi}(t)$ or
spinor twist number $N_{\Theta}(t)$ were ill-defined or
noninteger at time slice $t\in {\mathrm T}_{if}\equiv\{t_{i},t_{f}\}$.
In order to simplify the discussion, we exclude static field configurations
from our considerations.

The gauge field winding number $N_{\chi}(t)$, in particular,  is not well-defined
for a time slice $t=t^{(1)}\in {\mathrm
T}_{if}\,$ if the function $\chi(t^{(1)},r)$ has a zero, see
Eqs. (\ref{chi-polar}) and (\ref{windingN}).
Assume that the zero of $\chi$ occurs
for $t^{(1)}=t_i$ and that this is the only problem.
In this case, one can simply choose a real number $\delta t_i^{(1)}$, so
that the field $\chi(t,r)$ has no
zero at the time slice $t=t_{i}^{(1)}\equiv t_{i}+\delta t_i^{(1)}$.
For the new time interval
$[\,t_{i}^{(1)},t_{f}\,]$, one then obtains an integer-valued winding
factor $\Delta N_{\chi}[\,t_{f}, t_{i}^{(1)} \,]$.
The other case of having the zero of $\chi$ at $t^{(1)}=t_f$ can be treated
in the same way.

Alternatively, the spinor twist number $N_{\Theta}(t)$ can take a half-odd-integer
value for time slice $t=t^{(2)}\in {\mathrm T}_{if}$. Now recall that a
half-odd-integer spinor twist number implies the existence of a fermion
zero mode; see the paragraph below Eq. (\ref{UN}).
This makes it impossible to properly define the spectral flow for
the exact time interval $[\,t_i,t_f\,]$.  Assume that the zero mode occurs
for $t^{(2)}=t_i$ and that this is the only problem.
In this case, one can choose a
real number $\delta t_i^{(2)}$, so that $N_{\Theta}(t)$ takes a well-defined integer
value at the time slice $t=t_{i}^{(2)}\equiv t_{i}+\delta t_i^{(2)}$. This is
always possible, because a fermion zero mode
corresponds to an asymptotically \emph{unstable} solution $\Theta$ of
Eq. (\ref{FZMeq-Theta}).
For the new time interval $[\,t_{i}^{(2)},t_{f}\,]$, one
then obtains an integer-valued twist factor
$\Delta N_{\Theta}[\, t_{f}, t_{i}^{(2)} \,]$. The other case of having the
zero mode at $t^{(2)}=t_f$ can be treated in the same way.

Henceforth, we assume generic  spherically symmetric  $SU(2)$
gauge field backgrounds, so that the \rhs~of Eq.
(\ref{SelectionRule}) is well-defined and the sum of two integers.
It is, furthermore, clear that the winding factor
$\Delta N_{\chi}[\,t_f,t_i\,]$ is entirely determined by the background gauge
fields [see Eqs. (\ref{chi-polar}) and (\ref{windingN})].
But also the twist factor $\Delta N_{\Theta}[\,t_f,t_i\,]$
can be expressed solely in terms of background gauge
fields [see Eq. (\ref{twistfactorformal})].
The relation (\ref{SelectionRule}) thus connects a property of
the fermions, the spectral flow ${\cal F}$, to two characteristics of a
generic spherically symmetric $SU(2)$ gauge field background, the winding and
twist factors. This is the main result of the present paper.

\section{Spectral flow for L\"uscher-Schechter $SU(2)$ gauge fields}

In this section, we discuss the existence of fermion zero modes  and the
corresponding spectral flow for certain explicitly known
time-dependent spherically symmetric
solutions of the $SU(2)$ Yang--Mills equations. This allows for a nontrivial
check of
the relation (\ref{SelectionRule}) found in Sec. \ref{sec:LSspectralflow}.
Throughout this section and in the figures,
the same (arbitrary) mass scale is used to make the
spacetime coordinates and energy dimensionless.

\subsection{Brief review of the LS solutions}

The  solutions considered in this section are spherically symmetric solutions of
the $SU(2)$ gauge field equations, which
describe collapsing and re-expanding shells of energy.
The corresponding (1+1)-dimensional
field equations from the reduced action (\ref{SG-2d}) read
\begin{mathletters} \label{field-eq2}
\begin{eqnarray}
& &-\partial^{\mu}(r^{2}f_{\mu\nu})=2\,\mbox{Im}(\chi^{*}D_{\nu}\chi),\\
& &\left[-D^{2}+(|\,\chi |^{2}-1)/r^2 \right]\chi =0.
\end{eqnarray}
\end{mathletters}
Remarkably, L\"uscher and Schechter were able to obtain analytic solutions
of these coupled partial \diffeqs~\cite{L77,S77}.

The \LS~(LS) solutions can be represented as follows
(see Refs. \cite{FKS93,K95} and references therein):
\begin{mathletters} \label{Ansatz-LS}
\begin{eqnarray}
a_{\mu}&=&-q(\tau )\,\partial_{\mu}w ,\\
\alpha  &\equiv& \mbox{Re}\,\chi = 1+q(\tau )\,\cos^{2}w ,\\
\beta   &\equiv& \mbox{Im}\,\chi = (1/2)\,q(\tau )\,\sin 2w ,
\end{eqnarray}
\end{mathletters}
with the new coordinates
\begin{mathletters} \label{tau-w}
\begin{eqnarray}
\tau &\equiv& \mbox{sgn}(t)\:
\arccos\left(\frac{1+r^2-t^2}{\sqrt{(1+t^2-r^2)^2+4r^2}} \right),\\[0.2cm]
w &\equiv& \arctan\left(\frac{1-r^2+t^2}{2r}\right)\mbox{.}
\end{eqnarray}
\end{mathletters}
Using the \emph{Ansatz} (\ref{Ansatz-LS}),
the field equations (\ref{field-eq2}) are reduced
to a single nonlinear second-order \diffeq~for
$q(\tau )$,
\beq
\frac{\rmd^2 q}{\rmd\tau^2}+2\,q\,(q+1)\,(q+2)=0\mbox{.}
\label{eq-q}
\eeq

The ordinary differential equation (\ref{eq-q}) can be interpreted as
belonging to a mechanical system consisting of a particle
trapped in a double-well potential $V(q)\equiv \frac{1}{2}q^2(q+2)^2$. The
conserved total energy $\epsilon$ of the particle trapped in the potential $V$
is then
\beq
\epsilon = \frac{1}{2}\,\left(\frac{\rmd q}{\rmd\tau} \right)^2+V(q),\quad
V(q)\equiv q^2(q+2)^2/2 \mbox{.}
\label{double-well}
\eeq
The general solution of Eq. (\ref{eq-q}) depends on the energy parameter
$\epsilon$ and the $\tau$-translation parameter $\tau_{0}$, together
with a further discrete parameter $\zeta=\pm \,1$.
The solutions of Eq. (\ref{eq-q}) can be divided into two classes, one
with energy $\epsilon\leq 1/2$ and the other with energy
$\epsilon >1/2$. Explicitly, the LS solutions are \cite{L77,S77}
\begin{mathletters}
\begin{eqnarray}
q(\tau)&=&-1 + \zeta\,
(1+\sqrt{2\epsilon})^{1/2}\mbox{dn}\left[(1+\sqrt{2\epsilon})^{1/2}
(\tau -\tau_{0})|\,m^{-1}\right] \quad \mbox{for $\epsilon \leq 1/2$,}
\label{q-leq-half}\\
q(\tau )&=&-1+\zeta\,
(1+\sqrt{2\epsilon})^{1/2}\mbox{cn}\left[(8\epsilon
)^{1/4}(\tau
-\tau_{0})|\,m\right] \quad
\mbox{for $\epsilon > 1/2$,}\label{q-geq-half}
\end{eqnarray}
\end{mathletters}
\noindent with the modulus defined by
\beq \label{m}
  m\equiv\frac{1+ \sqrt{2\epsilon}}{2\sqrt{2\epsilon}}\mbox{.}
\eeq
Here, $\mbox{dn}[u|m]$ and $\mbox{cn}[u|m]$ are Jacobi elliptic functions
\cite{MF53,Math}.

For $\epsilon<1/2$, there exists no spacetime point where $\chi(t,r)$ vanishes.
For $\epsilon\geq 1/2$, on the other hand, there are zeros of
$\chi(t,r)$ at \cite{K95}
\beq
t_{n}=\tan\left(\tau_{0}+\frac{1+2n}{(8\epsilon )^{1/4}}\,K(m)\right), \:
\: r_{n}=\sqrt{1+t_{n}^2}\,,
\label{tn-rn}
\eeq
where $n$ is an integer that satisfies the condition
\beq
-\frac{\pi}{2}\leq\left(\tau_{0}+\frac{1+2n}{(8\epsilon)^{1/4}}\,K(m)\right)
\leq\frac{\pi}{2}\mbox{.}
\label{range-n}
\eeq
Here, $K(m)$ is the complete elliptic integral of the first kind,
\beq
  K(m)\equiv            \int_{0}^{1} \rmd u \,\left[(1-u^2)(1-mu^2)\,\right]^{-1/2}
\label{ellipticK}\mbox{.}
\eeq

The existence of a spacetime point $(t^*,r^*)$ where $\chi(t,r)$ vanishes is,
in general,
associated with the change of the winding number $N_{\chi}(t)$. It has been
shown in Ref. \cite{K95} that the change of the winding number $N_{\chi}(t)$ plays
an important role in fermion number violation (see also Ref. \cite{G94} for
related results). Indeed, we have studied several
LS solutions with $\epsilon<1/2$ and found that the necessary condition
(\ref{Kmax}) for the existence of a fermion zero mode is never satisfied.
In the following, we shall therefore only consider LS solutions with
$\epsilon\geq 1/2$. But before we turn to the fermion zero modes, we mention one
particular aspect of the LS gauge field background for $\epsilon\geq 1/2$.

\subsection{LS quasi-sphaleron}

For energy parameter $\epsilon\geq 1/2$, the field $\chi(t,r)$ has at least one
zero at a particular spacetime point. In order to simplify the analysis,
we take for the $\zeta$ and $\tau_0$ parameters in the solution (\ref{q-geq-half})
the following values:
\beq
\zeta=+1, \quad\tau_{0}=-(8\epsilon )^{-1/4}\,K(m),
\label{tau0}
\eeq
with $m$ defined by Eq. (\ref{m}).
For the choice of $\tau_{0}$  from Eq. (\ref{tau0}),
one of the zeros of $\chi(t,r)$ occurs at the time slice $t=t_{0}=0$.
Also note that for this $\tau_0$ the LS solution has
time-reversal (anti)symmetry, namely
$\rho (-t,r)=\rho (t,r)$ and ${\cal R}(-t,r)=-{\cal R}(t,r)$,
with $\rho (t,r)$ and ${\cal R}(t,r)$ defined in Eq. (\ref{DRdef}).
In the following, we consider the $t=0$ time slice of these particular
LS background gauge fields
[parameters $\zeta=+1$, $\tau_{0}$  from Eq. (\ref{tau0}), and  $\epsilon \geq 1/2$],
for which the zero of $\chi(0,r)$ occurs at $r=1$, according to Eq. (\ref{tn-rn}).

The LS gauge field background at $t=0$ is represented by
\beq
a_{\mu}(0,r)=0, \quad \chi (0,r)=\sin w (0,r),
\label{quasi-S}
\eeq
up to a $U(1)$ gauge transformation (\ref{gaugeT-2d}). Given that the real
function $\chi(0,r)$ vanishes
and changes sign at $r=1$, the configuration of Eq. (\ref{quasi-S})
qualitatively resembles the sphaleron solution of the \ew~$SU(2)$
Yang--Mills--Higgs theory \cite{KM84a,Y89}. For this reason, we call the configuration
given by Eq. (\ref{quasi-S}) the ``LS quasi-sphaleron.'' Note that the LS quasi-sphaleron
does not satisfy the static field equations, since the energy changes under a
scale transformation of the fields
(there is no natural mass scale for classical Yang--Mills theory).

We will now show that this LS quasi-sphaleron corresponds to the top
of a potential energy barrier
which separates configurations with different $N_{\chi}$, just like
the \ew~\sph~\cite{KM84a}. The energy functional \cite{FKS93,K95} for (1+1)-dimensional
gauge field solutions can, in fact, be written as
\beqa \label{ET}
E &=&E_{K}(t)+E_{P}(t),
\eeqa
with
\begin{mathletters}\label{EKP}
\begin{eqnarray}
E_{K}(t)&=&\frac{8\pi}{g^2}\int_{0}^{\infty}{\mathrm d}r \:
\left[\frac{1}{8\rho^2}
\left(\partial_{t}\rho^2\right)^2+\frac{1}{2\rho^2}
\left(\partial_{t}\psi\right)^2+\frac{1}{2\rho^2}
\left(\partial_{r}\psi\right)^2+\frac{\psi^2}{r^2}\right],\label{EK}\\[0.2cm]
E_{P}(t)&=&\frac{8\pi}{g^2}\int_{0}^{\infty}{\mathrm d}r \:
\left[\frac{1}{8\rho^2} \left(\partial_{r}\rho^2\right)^2+\frac{(\rho^2
-1)^2}{4r^2}\right],
\label{EP}
\end{eqnarray}
\end{mathletters}
\noindent where $\rho(t,r)$ equals $|\,\chi(t,r)|$ and
$\psi(t,r)$ is the (bosonic) gauge-invariant field defined by
\beq
-2\,\epsilon_{\mu\nu}\,\psi (t,r)\equiv r^2 f_{\mu\nu}(t,r)\mbox{.}\label{def-psi}
\eeq
Here, we have divided the energy into the kinetic part
$E_{K}$ and potential part $E_{P}$. The reason for putting the
$\psi$-dependent terms into the kinetic part of the energy is that,
for the gauge choice $a_{0}=0$, the scalar field $\psi$ becomes proportional
to the
time derivative of the $a_{1}$ field, namely $\psi =-r^2 \partial_{0}a_{1}$.

Consider then the potential energy $E_{P}(t)$ associated with our particular LS gauge
field solutions. Using the basic properties \cite{MF53} of the Jacobi elliptic function
$\mbox{cn}[u|m]$, one can prove that the potential energy
$E_{P}(t)$ has a local maximum at $t=0$, that is, for the LS quasi-sphaleron
configuration. Furthermore, the conserved total
energy of the LS solution with $\tau_0$ parameter (\ref{tau0}) is given by
\beq
E = 2\,\epsilon \, E_{\mathrm quasi-sph},
\eeq
in terms of the energy parameter $\epsilon$ and the static energy of the LS
quasi-sphaleron, $ E_{\mathrm quasi-sph}$ $\equiv$ $E_{P}(0)$.

\begin{figure}
\centerline{\psfig{figure=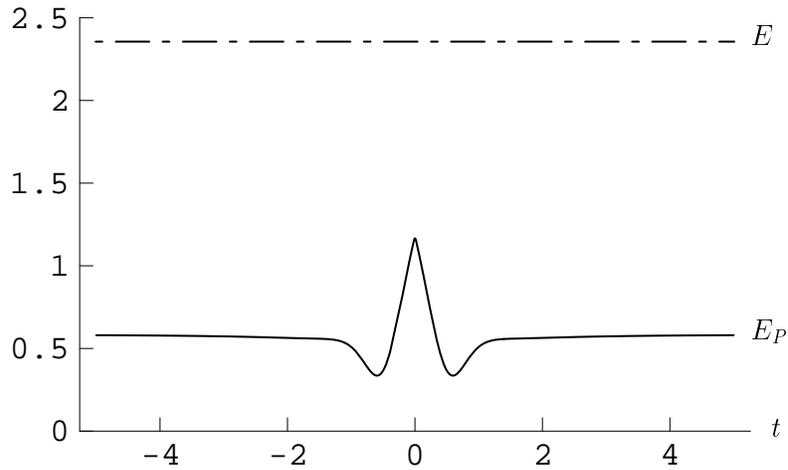,width=0.8\hsize}}\vspace{-3cm}
\caption{Time-development of the total energy $E$ and
potential energy $E_P (t)$ for the \LS~(LS) gauge field solution
(\ref{Ansatz-gauge}), (\ref{Ansatz-LS}), (\ref{q-geq-half})
with parameters  $\epsilon =1$, $\zeta=+1$, and $\tau_0\approx -1.4271$ from Eq.
(\ref{tau0}).
The potential energy $E_P (t)$, given by Eq. (\ref{EP}), is differentiable
for all times and  has a global maximum at
$t=0$. The configuration at $t=0$ is called the LS quasi-sphaleron.
}
\label{LSeps1}
\end{figure}

Figure \ref{LSeps1} shows the time development of $E_{P}(t)$
for the LS background with $\epsilon =1$, $\zeta=+1$, and $\tau_0 \approx -1.4271$ from
Eq. (\ref{tau0}). The corresponding topological charge $Q\approx -0.70$ is noninteger,
see Ref. \cite{FKS93} for further details. More importantly,
the potential barrier of Fig. \ref{LSeps1}  separates two regions with different
winding number  ($\Delta N_{\chi}=-1$).
The LS quasi-sphaleron at $t=0$ resembles in this respect also the
sphaleron of the \ewsm~\cite{KM84a}.

\begin{figure}
\centerline{\psfig{figure=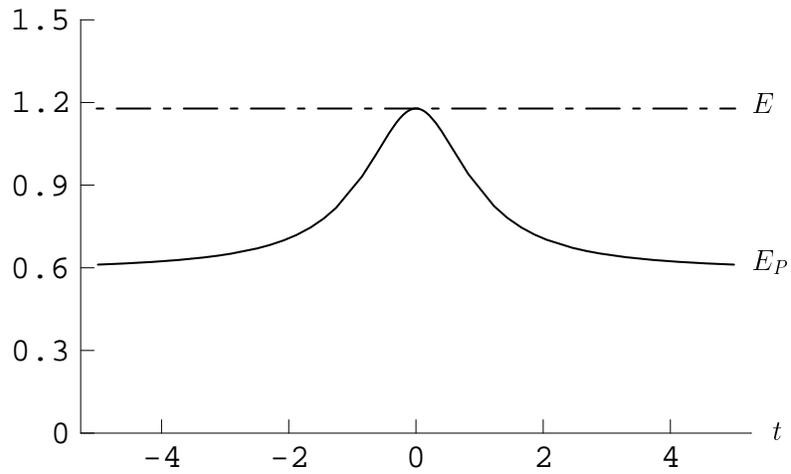,width=0.8\hsize}}\vspace{-3cm}
\caption{Time-development of the total energy $E$ and potential energy
$E_P (t)$ for the \AFF~(AFF) gauge field solution
(\ref{Ansatz-gauge}), (\ref{Ansatz-LS}), (\ref{AFFsolution}).
The AFF  configuration at $t=0$ coincides with the LS quasi-sphaleron
at $t=0$ in Fig. \ref{LSeps1}.
}
\label{AFF}
\end{figure}

For comparison, consider the LS gauge field background given by the trivial
solution $q=-1$ of Eq. (\ref{eq-q}), with the (1+1)-dimensional field
configurations
\beq
\psi(t,r) =0, \quad \chi(t,r) =\sin w(t,r)  \mbox{,}
\label{AFFsolution}
\eeq
as follows from Eqs. (\ref{gaugeT-2d}), (\ref{Ansatz-LS}), and (\ref{def-psi}).
This corresponds to the \AFF~(AFF) solution
\cite{AFF76}.
Note that the AFF gauge field coincides with the LS quasi-sphaleron
(\ref{quasi-S}) at $t=0$. On
the other hand, the AFF solution has complete time-reversal symmetry and the
kinetic energy $E_{K}(t)$ as given by Eq. (\ref{EK}) is zero at $t=0$ (see Fig.
\ref{AFF}). This result suggests that the
AFF solution provides the time-dependent gauge field
solution with minimum total
energy to form the LS quasi-sphaleron. In other words, the AFF gauge field
simulates an ``imploding and exploding LS quasi-sphaleron;''
cf. Ref. \cite{GS95}.
(For the \ew~\sph, the dynamics of the gauge and Higgs fields has been
studied numerically. See, for example, Refs. \cite{A91,RS96a}.)

\subsection{Fermion zero mode of the LS quasi-sphaleron}

We now turn to the fermion \zeieq~(\ref{FZMeq-D}) for the $t=0$
LS gauge field of the previous subsection, i.e. the LS quasi-sphaleron
(\ref{quasi-S}).
Note that the gauge-invariant function $\lambda(0,r)\equiv\rho(0,r) /r$ is
nondifferentiable at $r=1$, which is the only point
where $\rho(0,r) \equiv |\,\chi(0,r) |$ vanishes. For this reason, we introduce a
differentiable  field $\tilde{\lambda}(0,r)$, defined by
\begin{mathletters} \label{def-lambda-LS}
\beqa
\tilde{\lambda}(0,r)&\equiv& \kappa(0,r)/r, \\
\kappa(0,r)         &\equiv& \mbox{sgn}(1-r)\,|\,\chi(0,r)|\mbox{.}
\eeqa
\end{mathletters}
Then, $\chi(0,r)$ can be represented by
\beq \label{kappavarphi}
\chi (0,r)=\kappa (0,r)\,\exp[\,i\,\tilde{\varphi}(0,r)\,],
\eeq
where $\tilde{\varphi}(0,r)$ is a differentiable function of $r$.

It is a simple exercise to verify that the  LS field
$\kappa(0,r)$ has the following inversion symmetry:
\beq
\kappa (0,1/r)=-\kappa (0,r)\mbox{.}
\label{inverse-kappa}
\eeq
This inversion symmetry, most likely,
traces back to the conformal symmetry transformation
$x^{\mu}\rightarrow x^{\mu}/x^2$ of classical Yang--Mills theory.
Without loss of generality, we consider, in the following, smooth background
fields with $\kappa (0,0)=1$ and $\tilde{\varphi} (0,0)=0$.

In terms of the differentiable  fields $\tilde{\lambda}$
and $\tilde{\cal R}\equiv (a_{1}-\partial_{r}\tilde{\varphi})/2$, one
finds that the \zenfeqs~(\ref{FZMeq}) at $t=0$ become
\begin{mathletters}\label{FZMeq-LS}
\begin{eqnarray}
\partial_{r}\Theta &=&-\tilde{\lambda}\,\sin 2\Theta +\tilde{\cal R},
\label{FZMeq-Theta-LS}\\
\partial_{r}|\Psi |&=&\tilde{\lambda}\, |\Psi |\,\cos 2\Theta,
\label{FZMeq-Psi-LS}
\end{eqnarray}
\end{mathletters}
with boundary conditions
\beq \label{FZMeq-bcs-LS}
\Theta (0,0)=|\Psi (0,0)|=0\mbox{.}
\eeq
Using the inversion symmetry (\ref{inverse-kappa}) and the asymptotics
(\ref{local}) of the rotator, it follows from Eq. (\ref{FZMeq-Psi-LS})
that there exists a fermion zero mode if and
only if $\lim_{r\rightarrow \infty}\Theta(0,r) =N\pi$, with $N$ an integer.

We can now explicitly construct the fermion zero mode for the LS background at
$t=0$. Using the field equations \cite{FKS93,K95}, it is relatively straightforward
to show that
\beq \label{Rtilde}
\tilde{\cal R}(0,r)\propto \left.\frac{\rmd^2 q}{\rmd\tau^2}\right|_{\tau
=0}\propto\mbox{cn}[K(m)\,|\,m\,]=0,
\eeq
where the last identity can be found, for example,  in Ref. \cite{MF53}.
[Note that vanishing $\tilde{\cal R}$ does not contradict
the necessary condition (\ref{Kmax}) that was derived for Eqs. (\ref{FZMeq})
with a function $\lambda\geq 0$, whereas Eqs. (\ref{FZMeq-LS}) have a function
$\tilde{\lambda}$ that changes sign.]

For $\tilde{\cal R}=0$, the solutions to Eqs. (\ref{FZMeq-LS}) are simply given by
\begin{mathletters}
\label{FZM-LS-zero}
\beqa
\Theta (0,r)&=&0, \label{FZM-LS-zero-Theta}\\
 |\Psi (0,r)|&=&|\Psi (0,r_{0})|\, \exp\left[\,\int_{r_{0}}^{r}\rmd r'
\: \tilde{\lambda}(0,r')\right] \label{FZM-LS-zero-Psi}\mbox{.}
\eeqa
\end{mathletters}
The argument following Eq. (\ref{FZMeq-bcs-LS}) ensures that the solution
represented by Eqs. (\ref{FZM-LS-zero}ab) is normalizable, which completes the
construction of the fermion zero mode.
The inversion symmetry (\ref{inverse-kappa}), together with the result (\ref{Rtilde}),
provides a \emph{sufficient} condition for the existence of the fermion zero mode.

For $\epsilon\geq 1/2$, the fermion zero mode amplitude $|\Psi (0,r)|$ of Eq.
(\ref{FZM-LS-zero-Psi}) is shown in Fig. \ref{LSzmt0}, with an arbitrary normalization.
Specifically, the $(3+1)$-dimensional  fermion zero mode
is purely left-handed and given by Eqs. (\ref{Ansatz-fermion}) and (\ref{notation-2d}),
with the two-component spinor $\Psi_l=\Psi$ from
Eqs. (\ref{unitaryT}), (\ref{Psi-polar}), and (\ref{FZM-LS-zero}).
These last equations are to be evaluated with
the functions $\kappa(0,r)$ and $\tilde{\varphi}(0,r)$ defined in
Eq. (\ref{kappavarphi}). [Of course, the solution can also be
obtained from Eqs. (\ref{FZMeq}), which are given in terms of $\lambda(0,r)$ and
$\varphi(0,r)$.
In this case, the function $\Theta(0,r)$ is found to have a step function at
$r=1$, but so does the transformation matrix $\Lambda$ from Eq.
(\ref{Lambda}).]

To summarize, the LS quasi-sphaleron at $t=0$ has a chiral
fermion zero mode and resembles in this respect the \ew~\sph~which also has
a chiral fermion zero mode (see Refs.  \cite{N75,KM84b,R88} and references therein).
The fermion zero modes of  the LS quasi-sphaleron and the \ew~\sph~are
qualitatively the same. Moreover, there is spectral flow associated with both
the \ew~\sph~(see Refs. \cite{KM84b,R88,KB93}) and
the LS quasi-sphaleron (see Sec. V D 3 below). This behavior differs from
that of the AFF solution (an ``imploding and exploding LS quasi-sphaleron''),
for which the fermion zero mode exists at all
times \cite {GS95}, without level crossing.

\subsection{Level crossings for large energy parameter $\epsilon$}

\begin{figure}
\centerline{\psfig{figure=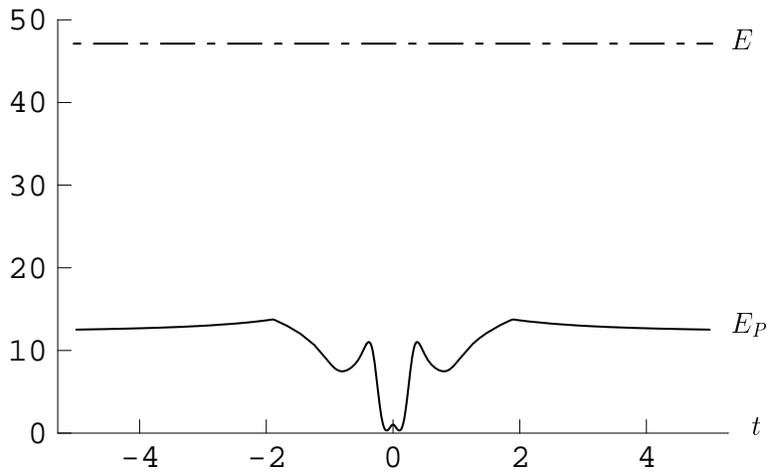,width=0.8\hsize}}\vspace{-3cm}
\caption{Time-development of the total energy $E$ and
potential energy $E_P (t)$ for the \LS~gauge field solution
(\ref{Ansatz-gauge}), (\ref{Ansatz-LS}), (\ref{q-geq-half})  with
parameters $\epsilon =20$, $\zeta=+1$, and $\tau_0\approx -0.54197$ from Eq.
(\ref{tau0}).
See Fig. \ref{LSeps20b} for a close-up of $E_P (t)$ near $t=0$.
}
\label{LSeps20a}
\end{figure}

\begin{figure}
\centerline{\psfig{figure=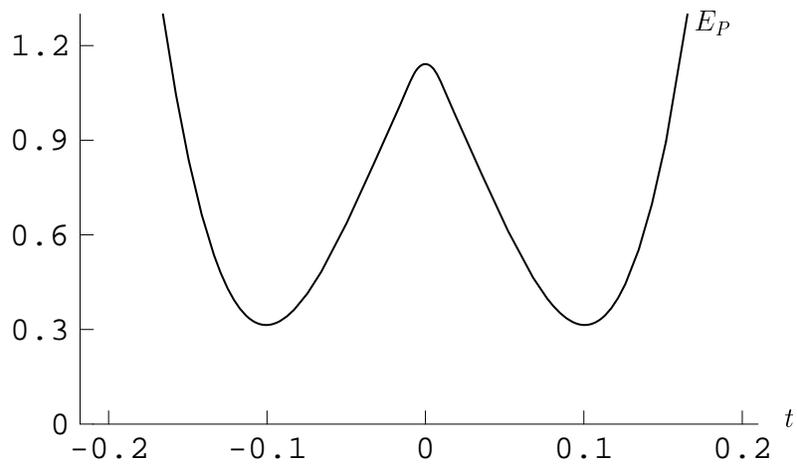,width=0.8\hsize}}\vspace{-3cm}
\caption{Same as  Fig. \ref{LSeps20a}.
Close-up of the potential energy $E_P (t)$ near $t=0$. The LS quasi-sphaleron
at $t=0$ is only a local maximum of $E_P (t)$, unlike
the case of Fig. \ref{LSeps1}.
}
\label{LSeps20b}
\end{figure}

Throughout this subsection, we consider the specific LS gauge field solution with
parameters $\epsilon =20$, $\zeta=+1$, and $\tau_{0}\approx -0.54197$ from
Eq. (\ref{tau0}). Figures \ref{LSeps20a} and \ref{LSeps20b} give
the behavior of the potential energy $E_P(t)$ of this solution.
The corresponding topological charge $Q\approx -0.13$ is noninteger; cf. Ref. \cite{FKS93}.
The LS quasi-sphaleron at $t=0$ is the same as for the $\epsilon =1$ case,
but no longer corresponds to a global maximum of $E_P(t)$.

In order to determine the spectral flow for this particular gauge field
background, \emph{all} fermion zero modes need to be determined, to which we turn
first.

\subsubsection{Extra fermion zero modes from changes in winding number}

According to Eq. (\ref{tn-rn}), there are zeros of this LS  field
$\chi(t,r)$ at the spacetime points
\beqa
(t_{-1},r_{-1})&\approx& (-1.889,2.137), \quad (t_{0},r_{0})= (0,1),\nonumber\\
(t_{+1},r_{+1})  &\approx& (+1.889,2.137)\mbox{.}
\label{tn-rn-20}
\eeqa
The winding number $N_{\chi}(t)$
takes the following values in the different time regions:
\beq
N_{\chi}(t)= \left\{ \begin{array}{ll}
-1& \quad {\mathrm for} \quad t\in (-\infty, t_{-1}),\\
0 & \quad {\mathrm for} \quad t\in (t_{-1},t_{0}), \\
-1& \quad {\mathrm for} \quad t\in (t_{0},t_{+1}),\\
0 & \quad {\mathrm for} \quad t\in (t_{+1},+\infty ).
\end{array}\right.
\eeq
This gives the global winding factor
\beq
\Delta N_{\chi}[ +\infty, -\infty \,]=0-(-1) =1\mbox{.}
\label{twistF-LS}
\eeq

\begin{figure}
\centerline{\psfig{figure=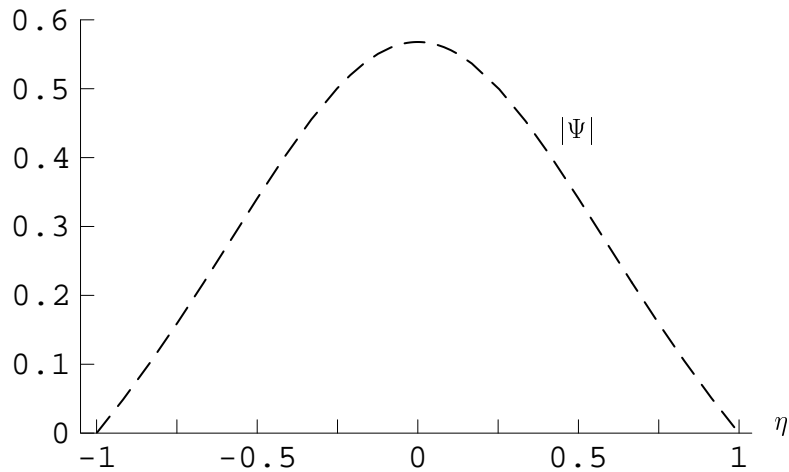,width=0.8\hsize}}\vspace{-3cm}
\caption{Profile function $|\Psi(0,r)|$
of the  fermion zero mode (\ref{FZM-LS-zero-Psi})
of the LS quasi-sphaleron, which corresponds to the
$t=0$ configuration of Fig. \ref{LSeps1}.
The dashed curve gives $|\Psi(0,r)|$  with an arbitrary normalization.
The inversion symmetry $r \rightarrow 1/r$ is made manifest by use of the
compact radial coordinate $\eta\equiv (r-1)/(r+1)$.
}
\label{LSzmt0}
\end{figure}

As discussed in Sec. V C, there exists a fermion zero mode at $t=t_0=0$ (see
Fig. \ref{LSzmt0}). But there are two more fermion zero modes precisely at
$t=t_{-1}$ and $t=t_{+1}$. Three preliminary steps are necessary for the proof.

First, define smooth fields $\tilde{\lambda}(t_n,r)$ and $\tilde{\cal R}(t_n,r)$
at the time slices $t=t_n$, for $n=\pm 1$,
\begin{mathletters} \label{def-tilde}
\beqa
\tilde{\lambda}(t_n,r)&\equiv& \kappa(t_n,r)/r\;,\\
\tilde{\cal R}(t_n,r)&\equiv& \left[\,a_1(t_n,r)
-\partial_r\,\tilde{\varphi}(t_n,r)\,\right]/2\;,
\eeqa
\end{mathletters}
with the differentiable function
$\kappa (t_n,r)$ $\equiv$ $\rho(t_n,r)$ $\times$ ${\mathrm sgn}(r_n-r)$
and the smooth
argument $\tilde{\varphi}(t_n,r)$ of the Higgs-like field
$\chi =\kappa\,\exp[\,i\,\tilde{\varphi}\,]$.

Second, perform a scale
transformation $x=r/r_{n}$, so that the radial point $r=r_{n}$
where the field $\chi (t_n,r)$ vanishes corresponds to $x=1$.

Third, establish that the {\LS} solution for the parameters chosen has the
following inversion symmetry at fixed time slices $t=t_n\,$:
\begin{mathletters} \label{inverse-LS}
\beqa
\tilde{\lambda}(t_n,1/x)&=&-x^2\,\tilde{\lambda}(t_n,x)\;,\\
\tilde{\cal R}(t_n,1/x) &=&-x^2 \,\tilde{\cal R} (t_n,x)\;,
\eeqa
\end{mathletters}
for $n$ $=$ $\pm\,1$.
The existence of this inversion symmetry has been verified
analytically with the help of M{\sc athematica} 4.0 \cite{Math}.

After these preliminaries, we turn to the possible  existence of fermion zero modes
at the time slices $t=t_{\pm 1}$. Consider the \zenfeqs~given by Eqs.
(\ref{FZMeq}) at $t=t_n$,  for $n=\pm\,1$, with
the smooth background fields $\tilde{\lambda}(t_n,r)$ and $\tilde{\cal R}(t_n,r)$.
The chiral \YMth~(\ref{action4d}) is scale invariant
and the rescaling $x=r/r_{\pm 1}$ at $t=t_{\pm 1}$ does not alter the
structure of Eqs. (\ref{FZMeq}):
\begin{mathletters}\label{FZMeq-inverse}
\beqa
\frac{\rmd \Theta}{\rmd x} &=&-\tilde{\lambda}(x)\,\sin 2\Theta(x)
+\tilde{{\cal R}}(x)\;, \quad \Theta (0)=0\;,\\[0.2cm]
\frac{\rmd |\Psi|}{\rmd x}&=&\tilde{\lambda}(x)\,|\Psi (x)|\,\cos 2\Theta (x)\;,
\quad\:\:\,\, |\Psi (0)|=0\;,
\eeqa
\end{mathletters}
with the dependence on $t_{\pm 1}$ temporarily dropped.
The \diffeqs~(\ref{FZMeq-inverse}) are symmetric under the inversion
transformation $x\leftrightarrow 1/x$ and so are their solutions $\Theta(x)$
and $|\Psi(x)|$. The inversion symmetry implies
\beq
\lim_{x\rightarrow 0}\Theta =\lim_{x\rightarrow \infty}\Theta=0\;, \quad
\lim_{x\rightarrow 0}|\Psi| =\lim_{x\rightarrow \infty}|\Psi|=0\;,
\label{FZM-at-tplusminus1}
\eeq
with $|\Psi|\propto 1/x$ for large $x$. This shows that there exist fermion
zero modes at both $t=t_{-1}$ and $t=t_{+1}$.

The inversion symmetry
(\ref{inverse-LS}), after the appropriate  scale transformation, provides
again a \emph{sufficient} condition for the existence of fermion zero modes
at $t=t_{\pm 1}$.
For {\LS} gauge field backgrounds with arbitrary $\tau_0$ and $\epsilon\geq 1/2$, the
inversion symmetry  (\ref{inverse-LS}) holds, in fact, at any time slice $t=t_n$ where
$\chi (t,r)$ has a zero. This then proves the existence of fermion zero modes
at all $t_n$.

\begin{figure}
\centerline{\psfig{figure=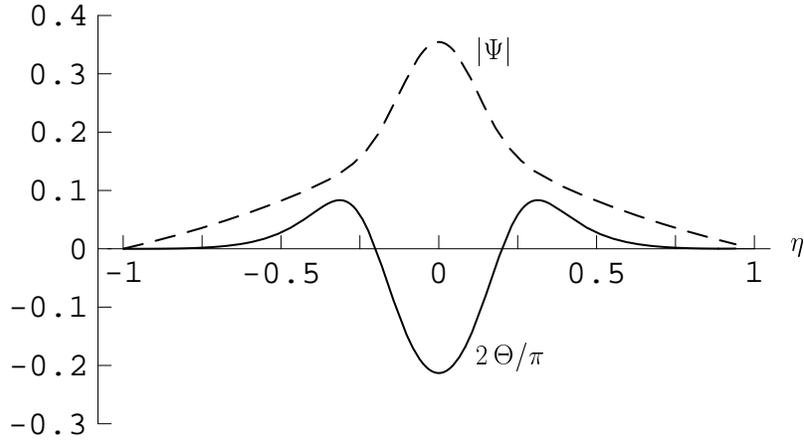,width=0.8\hsize}}\vspace{-3cm}
\caption{Numerical solutions for the profile functions
$\Theta(t_{-1},r)$ and $|\Psi(t_{-1},r)|$
of the  fermion zero mode at $t=t_{-1}\approx -1.889$  for the \LS~background
gauge field (\ref{Ansatz-gauge}), (\ref{Ansatz-LS}), (\ref{q-geq-half})
with parameters $\epsilon =20$, $\zeta=+1$, and $\tau_0\approx -0.54197$.
The solid curve corresponds to $\Theta(t_{-1},r) \times 2/\pi$ and the dashed
curve to $|\Psi(t_{-1},r)|$ with an arbitrary normalization.
The inversion symmetry $x \rightarrow 1/x$, with $x \equiv r/r_{-1}$,
is made manifest by use of the
compact radial coordinate $\eta\equiv (x-1)/(x+1)$.
}
\label{LSzmtminus1}
\end{figure}

Figure \ref{LSzmtminus1} gives the profile functions of the fermion zero mode
at $t=t_{-1}$, obtained from the numerical solution of Eqs. (\ref{FZMeq-inverse}).
The profile functions of the fermion zero mode at $t=t_{+1}$
are identical, except for a change of sign of $\Theta$.
These  functions are quantitatively different from those of the LS
quasi-\sph~[see Eqs. (\ref{FZM-LS-zero}) and Fig. \ref{LSzmt0}], but
qualitatively the same.

\subsubsection{Extra fermion zero modes from changes in twist number}

In order to locate all possible level crossings, we are guided by
the change of the spinor twist number $N_{\Theta}(t)$. The spinor twist number
$N_{\Theta}(t)$ takes the following values in the different time regions:
\beq
N_{\Theta}(t)= \left\{ \begin{array}{ll}
+1 & \quad {\mathrm for} \quad t\in (-\infty,-t_{a}),\\
0  & \quad {\mathrm for} \quad t\in (-t_{a},+t_{a}),\\
-1 & \quad {\mathrm for} \quad t\in (+t_{a},+\infty),
\end{array}\right.
\eeq
with the numerical estimate $t_{a}\approx 2.924$.
The corresponding twist factor is thus given by
\beq
\Delta N_{\Theta}[ +\infty, -\infty \,]=-1 - 1  =-2\mbox{.}
\label{twistF-e20}
\eeq

In addition to the fermion zero modes at $t_0$ and $t_{\pm 1}$, which
are associated with the change of the gauge field winding number  $N_{\chi}(t)$,
there exist two more fermion zero modes precisely at $t=\pm\,t_{a}$.
The analysis of these  fermion zero modes is straightforward
and the normalizability condition is found to hold, provided
$\Theta(\pm t_a,r)$ approaches a half-odd-integer multiple of $\pi$ as
$r\rightarrow \infty$; see Eq. (\ref{FZMeq-Psi}).
Note that the exact value of $t_a$ is defined implicitly by the relation
$\lim_{R\rightarrow \infty}$ $\Theta(\pm\,t_a,R)$ $=$ $\mp\,\pi/2$,
where the solution $\Theta(\pm\,t_a,R)$ of the \diffeq~(\ref{FZMeq-Theta})
with boundary condition (\ref{BCTheta}) can be obtained by the method of
successive approximations \cite{I56,H69}.

\begin{figure}
\centerline{\psfig{figure=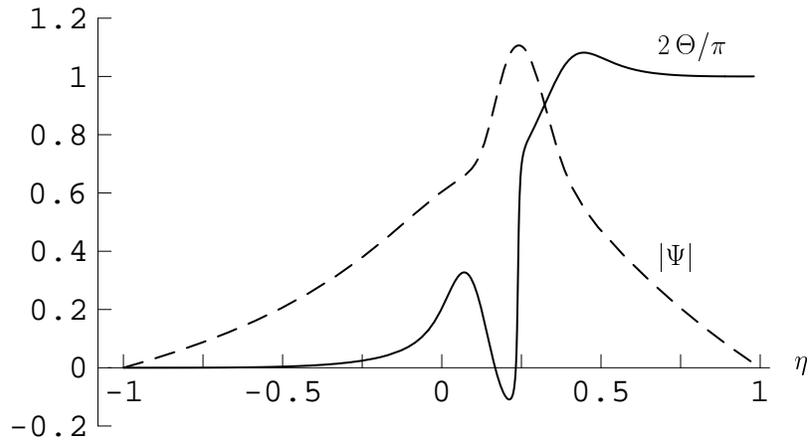,width=0.8\hsize}}\vspace{-3cm}
\caption{Same as Fig. \ref{LSzmtminus1}, but for the  fermion zero mode at
$t=- t_{a}\approx -2.924$.
The solid curve corresponds to $\Theta(-t_a,r) \times 2/\pi$ and the dashed
curve to $|\Psi(-t_a,r)|$ with an arbitrary normalization.
Both functions are plotted against the compact
radial coordinate $\eta\equiv (x-1)/(x+1)$, with $x \equiv r/2$.
}
\label{LSzmtminusa}
\end{figure}

\begin{figure}
\centerline{\psfig{figure=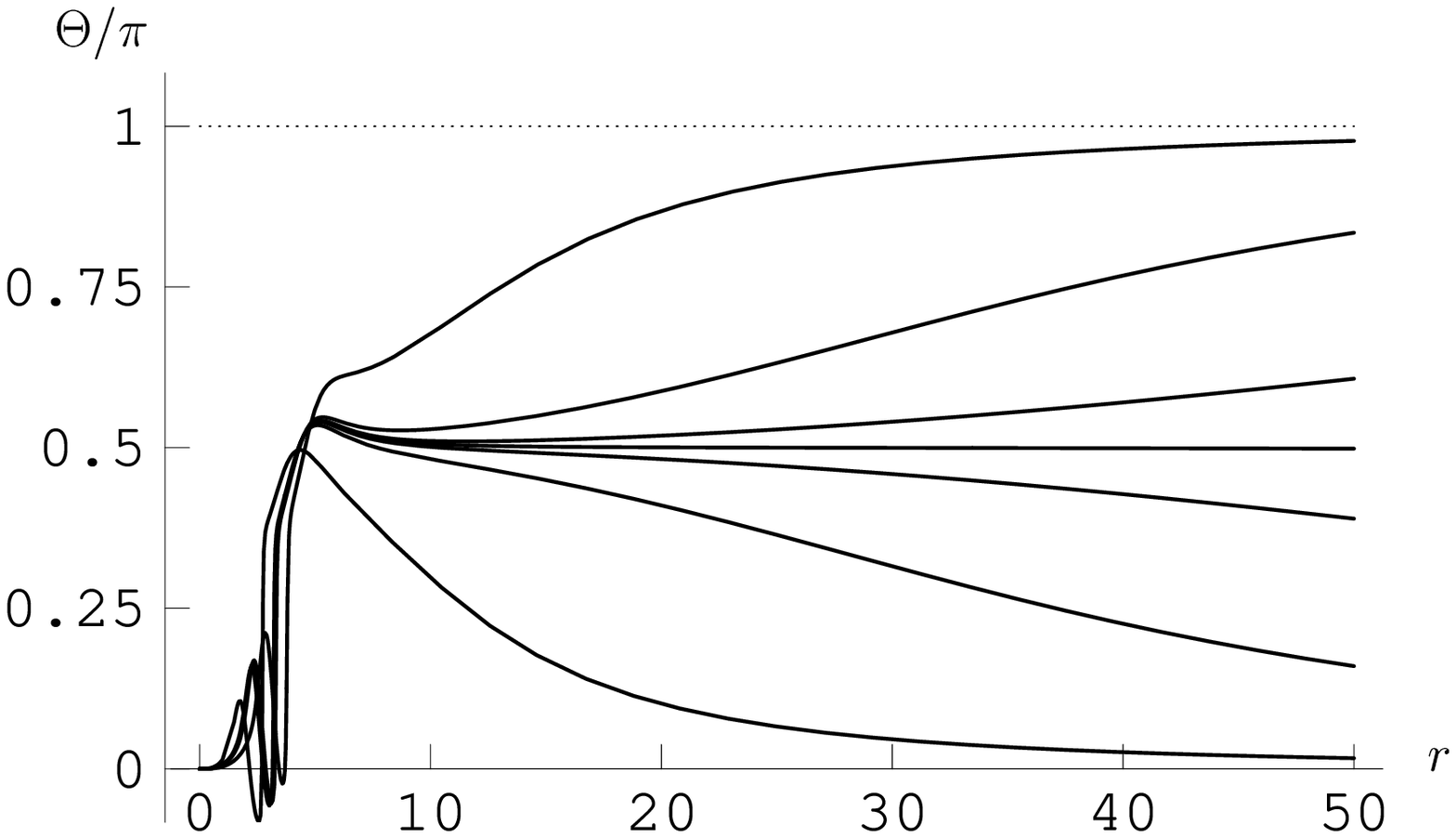,width=0.8\hsize}}\vspace{-3cm}
\caption{Numerical solutions $\Theta(t,r)$ of the \tReq~(\ref{FZMeq-Theta}),
with boundary condition $\Theta(t,0)$ $=$ $0$, at different times  around
$t=- t_{a}\approx -2.924$ for the \LS~background gauge field (\ref{Ansatz-gauge}),
(\ref{Ansatz-LS}), (\ref{q-geq-half})
with parameters $\epsilon =20$, $\zeta=+1$, and $\tau_0\approx -0.54197$.
For $t=- t_{a}$, there is a normalizable fermion zero mode,
with $\lim_{r\rightarrow \infty}\Theta(-t_a,r) = \pi/2$ (see Fig. \ref{LSzmtminusa}).
The different solid curves for $r$ $\geq$ $10$, from top to bottom, correspond to
$t+t_{a}$ $=$ $-0.50$, $-0.05$, $-0.01$, $0.00$, $0.01$, $0.05$, and $0.50$.
With increasing time, the $\Theta$ values at large $r$ move
\emph{toward} the constant $\Theta$ value at $r=0$.
}
\label{LSThetaminta}
\end{figure}

\begin{figure}
\centerline{\psfig{figure=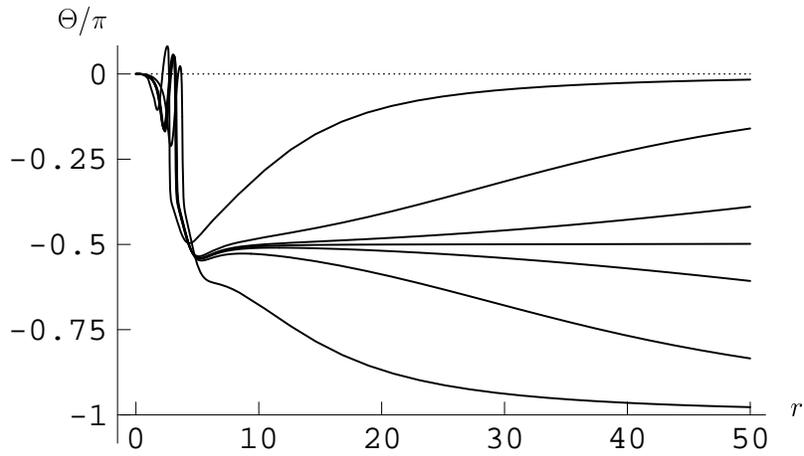,width=0.8\hsize}}\vspace{-3cm}
\caption{Same as Fig. \ref{LSThetaminta}, but for times around
$t$ $=$ $t_{a}$ $\approx$ $2.924$.
For $t=t_{a}$, there is a normalizable fermion zero mode,
with $\lim_{r\rightarrow \infty}\Theta(t_a,r) = -\pi/2$.
The different solid curves for $r$ $\geq$ $10$, from top to bottom, correspond to
$t-t_{a}$ $=$ $-0.50$, $-0.05$, $-0.01$, $0.00$, $0.01$, $0.05$, and $0.50$.
With increasing time, the $\Theta$ values at large $r$ move
\emph{away} from the constant $\Theta$ value at $r=0$.
}
\label{LSThetaplusta}
\end{figure}

Figure \ref{LSzmtminusa} gives the
profile functions of the fermion zero mode at $t=-t_{a}$, obtained from
the numerical solution of Eqs. (\ref{FZMeq}) and (\ref{FZMeq-bcs}).
The profile functions of the fermion zero mode at $t=+t_{a}$
are identical, except for a change of sign of $\Theta$.
Figures \ref{LSThetaminta} and \ref{LSThetaplusta} show
the time-variation of the solutions $\Theta(t,r)$ of the \tReq~(\ref{FZMeq-Theta})
around $t=\pm t_{a}$, which demonstrates that the fermion zero modes at $t=\pm t_{a}$
sit at bifurcation points for different $N_{\Theta}$'s.
[These results provide an example for the general discussion leading up to
Eqs. (\ref{twist-plusmin}) in Sec. IV A 2.] Obviously,
Figs. \ref{LSThetaminta} and \ref{LSThetaplusta} are related, because
of the ti\-me-re\-flec\-tion properties of the background fields mentioned below
Eq. (\ref{tau0}).

The fermion zero modes at $t=\pm t_a$ are qualitatively  different from the
ones at $t_0$ and $t_{\pm 1}$ (compare Fig. \ref{LSzmtminusa} with
Figs. \ref{LSzmt0} and \ref{LSzmtminus1}).
These fermion zero modes occur, in fact, for Higgs-like fields
$\chi(\pm t_a,r)$ without zeros. This differs from the cases discussed in
the literature \cite{K95,G94,W81}.
Apparently, the long-range behavior of the
background $SU(2)$ gauge fields plays a crucial role for the existence
of these extra fermion zero modes (see also the discussion in Sec. VI).

\subsubsection{Spectral flow} \label{sec:LSspectralflow}

For the fermion zero modes at $t=-t_{a}$, $t_{-1}$, $t_0$,
$t_{+1}$, and $+t_{a}$, we calculate the following
time-gradients of the energy eigenvalue of the effective Dirac Hamiltonian:
\begin{eqnarray}
\left. \dEdt \right|_{t=-t_{a}}&=&
\left. \dEdt \right|_{t=+t_{a}}  \approx
-0.03<0,\nonumber\\[0.2cm]
\left. \dEdt \right|_{t=t_{-1}}&=&
\left. \dEdt \right|_{t=t_{+1}}\:\approx
+0.08>0,\nonumber\\[0.2cm]
\left. \dEdt \right|_{t=t_{0}}&\approx&
-5.00<0\mbox{.}
\label{LevelCrossing-LS}
\end{eqnarray}
In addition, we have checked that there are no further fermion zero modes, at least
for the time interval $[-200,+200]$.

The level crossings corresponding to Eq. (\ref{LevelCrossing-LS})
give the following value
for the spectral flow (starting from $t=-t_{a}$ and ending at $t=+t_{a}$):
\beq
{\cal F}[ +\infty, -\infty \,]=-1+1-1+1-1=-1\mbox{.}\label{Spectral-crossing}
\eeq
It would certainly be interesting to calculate the spectrum of the time-dependent
effective Dirac Hamiltonian numerically (cf. Ref. \cite{A91})
in order to see whether or not the pattern
(\ref{Spectral-crossing}) corresponds to a \emph{single} energy level crossing $E=0$
five times.
Anyway, the total value (\ref{Spectral-crossing}) agrees with the spectral flow
obtained from the relation (\ref{SelectionRule}):
\beqa
{\cal F}[ +\infty, -\infty \,]&=& \Delta N_{\chi}[ +\infty, -\infty \,] +
                                  \Delta N_{\Theta}[ +\infty, -\infty \,]\nonumber\\
                              &=&1-2=-1\mbox{,}
\label{Spectral-SelectionRule}
\eeqa
where the results (\ref{twistF-LS}) and (\ref{twistF-e20}) have been used.
This demonstrates the role of the twist factor $\Delta N_{\Theta}$ for the
spectral flow, which has not been noticed before to our knowledge.

Recall that the explicit results of the present subsection are for the
particular LS gauge field background with energy parameter $\epsilon=20$,
together with $\zeta$ and $\tau_0$ from Eq. (\ref{tau0}).
The spectral flow for LS background gauge fields turns out to be solely given by the
winding factor  $\Delta N_{\chi}$ if $\epsilon$ is smaller than
$\epsilon^{*}\approx 5.37071$,
which is the numerical solution of the following equation:
\beq
\epsilon^{*}=\frac{1}{8}\left[\frac{4}{\pi}
K\left(\frac{1+\sqrt{2\epsilon^{*}}}{2\sqrt{2\epsilon^{*}}}\right)\right]^4
\mbox{.}
\eeq
Briefly, the argument runs as follows. First, the numerical results (and a
heuristic argument) give $t_a \geq t_{+1}$. Second, the times $t_{\pm 1}$
move toward $\pm\infty$ as $\epsilon$ approaches $\epsilon^{*}$ from above;
see Eqs. (\ref{tn-rn}) and (\ref{tau0}).
Together, this implies
that the contribution of the twist factor to the spectral flow
vanishes for $\epsilon \leq \epsilon^{*}$,
giving ${\cal F}[ +\infty, -\infty \,]=\Delta N_{\chi}[ +\infty, -\infty \,]$.
(The winding factor  $\Delta N_{\chi}[ +\infty, -\infty \,]$ is,
of course, identically zero for $\epsilon <1/2$.)

To summarize, the twist factor $\Delta N_{\Theta}$
can play a significant role for the spectral flow in certain LS
gauge field backgrounds, provided the  energy parameter
is large enough ($\epsilon$ $>$ $\epsilon^{*}$).

\section{Discussion}

In this paper, we have studied real-time anomalous fermion number violation by
directly investigating the zero-eigenvalue equation (\ref{Hamiltonianeq})
of the time-dependent  effective  Dirac
Hamiltonian for spherically symmetric massless chiral fermions and
$SU(2)$ Yang--Mills gauge field backgrounds. For these spherically symmetric
$SU(2)$ gauge field backgrounds, we have found a relation between the spectral
flow and two characteristics of the gauge fields. Physics applications
of this result are based on the assumption that anomalous production of
fermions is confined to the spherically symmetric partial wave; cf. Ref.
\cite{Y89}. Perhaps the most important application would be for
\ew~baryon number violation in the early universe, in particular at
temperatures above the \ew~phase transition (see Ref. \cite{RS96b} for a review).

Since we adopt an approach different from the one of previous work
\cite{C80,GH95,K95},
we are able to observe certain new features of real-time fermion number violation
within the spherically symmetric \emph{Ansatz}.
These features include the spinor twist number $N_\Theta(t)$ obtained from the Riccati
equation (\ref{Riccati}) at a single fixed time $t$ and the
corresponding twist factor $\Delta N_\Theta[\,t_f,t_i\,]$
which is the change of the spinor twist
number over the time interval $[\,t_i,t_f\,]$, with $t_i$ $<$ $t_f$.
Relation (\ref{SelectionRule}) then gives the spectral flow ${\cal F}$ as the sum
of this twist factor $\Delta N_\Theta$ and the winding factor
$\Delta N_\chi$, which is the change of the Chern-Simons number $N_\chi$ of the
associated vacuum sectors of the gauge field background.
Mathematically, the relation (\ref{SelectionRule}) takes the form of an index
theorem, restricted to  spherically symmetric fields; cf. Ref. \cite{C80}.

In order to get better insight into the meaning of this relation (\ref{SelectionRule}),
we have investigated level crossings for a particular class of \LS~(LS)
gauge field solutions \cite{L77,S77}. The results clearly demonstrate
the role of the twist factor for the spectral flow. See, in particular, Eqs.
(\ref{Spectral-crossing}) and (\ref{Spectral-SelectionRule}). The nonvanishing global
effect of the twist factor on the spectral flow is partly due to the fact that the
LS gauge fields form propagating \emph{solitons} in the effective
(1+1)-dimensional theory; cf. Refs. \cite{FKS93,K95}.
The fields are thus nondissipative in the
(1+1)-dimensional world. For such background gauge fields, the field-theoretic
approach adopted in Refs. \cite{GH95,K95} is, strictly speaking, not applicable.

Although nondissipative in the (1+1)-dimensional context, these LS gauge
field solutions are dissipative in (3+1)-dimensional
spacetime. The (3+1)-dimensional energy density, which is
obtained from the (1+1)-dimensional energy density (\ref{EKP})
by dividing by $4\pi r^2$, approaches zero uniformly for
early and late times ($t$ $\rightarrow$ $\pm\infty$).

At this moment, we propose to classify dissipative spherically symmetric
$SU(2)$ gauge field  solutions into two categories.  A spherically
symmetric $SU(2)$ gauge field  solution is called \emph{strongly dissipative},
if both the (3+1)-dimensional and (1+1)-dimensional energy densities
approach zero uniformly for large times ($t$ $\rightarrow$ $\pm\infty$).
On the other hand, a spherically symmetric $SU(2)$ gauge field  solution is called
\emph{weakly dissipative},
if the (3+1)-dimensional energy density dissipates with time, but not
the (1+1)-dimensional energy density.  Note that the
LS gauge field  solutions considered in Sec. V D
are weakly dissipative, according to this terminology.

For strongly dissipative spherically symmetric $SU(2)$ gauge field solutions, the
rotator ${\cal R}(t,r)$ in the \tReq~(\ref{FZMeq-Theta})
lacks the strength to give a nonzero
spinor twist number $N_{\Theta}$ at large times (see Appendix A for the proof).
The relation (\ref{SelectionRule}) then predicts that the spectral flow
${\cal F}$ is solely given by the winding factor $\Delta N_{\chi}$,
which reproduces the known result \cite{GH95,K95}.
An isolated change of spinor twist number can still
contribute to the \emph{local} pattern of level crossing
(\ref{spectral-local}).

For  weakly dissipative or nondissipative spherically symmetric $SU(2)$
gauge field solutions, a nonvanishing twist factor
$\Delta N_{\Theta}$ can even make a \emph{global}
contribution to the spectral flow ${\cal F}$. As mentioned above, this has been
verified for certain LS  solutions \cite{L77,S77}.
This behavior does not follow directly
from the perturbative triangle anomaly \cite{A69,BJ69,H76},
which detects only the (noninteger) topological charge.
Recall that the standard perturbative calculations
(Feynman diagrams) essentially neglect the interactions of incoming and outgoing
particles, i.e. the interactions are ``turned off'' in the asymptotic
regions \cite{F49}.

To summarize, the spectral flow result (\ref{SelectionRule}) does not assume
strongly dissipative spherically symmetric $SU(2)$ \YM~gauge field
backgrounds, in contrast to previous studies \cite{C80,GH95,K95}. It may,
therefore, be applied to real-time anomalous fermion number violation for weakly
dissipative or nondissipative  spherically symmetric $SU(2)$ \YM~gauge field
backgrounds. Moreover, preliminary results indicate that
the relation  (\ref{SelectionRule}) can be adapted to the case
of chiral fermions interacting with spherically symmetric $SU(2)$ Yang--Mills and
Higgs fields.

The main outstanding problem is, of course, to understand
the role of the (appropriately generalized) twist factor in the
\emph{full} (3+1)-dimensional $SU(2)$ Yang--Mills
theory, not just the subspace of spherically symmetric configurations.
Also, the corresponding index theorem needs to be established for the
long-range Yang--Mills
gauge fields considered; cf. Ref. \cite{W81}
and references therein. Finally, the proper definition (if at all possible)
of the second-quantized fermion number operator in general
nondissipative Yang--Mills gauge field backgrounds requires further study.

\begin{appendix}

\section{Spinor twist number and strongly dissipative $SU(2)$ gauge fields}

In this appendix, we calculate the asymptotic
spinor twist number $N_{\Theta}(t)$, defined by Eqs.
(\ref{FZMeq-Theta}), (\ref{BCTheta}), and (\ref{twistN}) in the main text,
for strongly dissipative spherically symmetric $SU(2)$ gauge field solutions.

 From the (1+1)-dimensional energy density (\ref{EKP}) and the
corresponding field equations (see, in particular Eqs. (3.8) and (3.10)
of Ref. \cite{K95}), the condition of strong dissipation implies that
there exists a small positive
quantity $\epsilon(t)\ll 1$ at large $|t|$ , so that
\beq
|\,r^{-1}-\lambda(t,r)|<\epsilon(t) \quad \mbox{and}
\quad |{\cal R}(t,r)|<\epsilon(t)\;,
\label{ineq:tlarge}
\eeq
for $0\leq r<\infty$, together with the limit
\beq
\lim_{|t|\rightarrow\infty}\epsilon(t)=0\;.
\eeq
See Eqs. (\ref{notation-2d}), (\ref{chi-polar}), and (\ref{DRdef})
in the main text for the definition of $\lambda(t,r)$ and
${\cal R}(t,r)$.

 From the bounds (\ref{ineq:tlarge}), we immediately obtain
\beq
0< r^{-1}-2\,\epsilon(t)  < \lambda(t,r)-|{\cal R}(t,r)|\;,
\label{ineq:rfinite}
\eeq
for $0\leq r< R_{\epsilon}\equiv [2\,\epsilon(t)]^{-1}$.
Since regular finite-energy gauge fields obey
$\lim_{r\rightarrow\infty}{\cal R}/\lambda =0$ [see Eq. (\ref{local}) in the
main text], we have from Eq. (\ref{ineq:rfinite}) the following inequality
for arbitrary $r$:
\beq
0 \leq |{\cal R}(t,r)| < \lambda (t,r) \;,
\label{ineq:lambdaMR}
\eeq
provided $|t|$ is sufficiently large.

With the inequality (\ref{ineq:lambdaMR}) in hand, we are able to
establish that the spinor twist number $N_{\Theta}(t)$ vanishes asymptotically.
This can be shown by contradiction. Consider the \tReq,
\begin{mathletters} \label{FZMeqtlarge}
\beqa
\partial_r \Theta (t,r)&=&{\cal D}(t,r) +{\cal R}(t,r),\ \quad \Theta (t,0)=0, \\
{\cal D}(t,r) &\equiv&-\lambda(t,r)\sin 2\Theta (t,r)\;,
\eeqa
\end{mathletters}
at sufficiently large $|t|$.
Now, assume that there exists a time slice
$t=\bar{t}$ where the background
fields obey inequality (\ref{ineq:lambdaMR}) and that the solution
$\Theta (\bar{t},r)$ of Eq. (\ref{FZMeqtlarge}) belongs to the class
$S_{N}(\bar{t})$, with \emph{positive} integer
$N$ (the case of negative $N$ will be dealt with later).
See Eq. (\ref{SN}) in the main text
for the definition of the solution class $S_N$.

Strong dissipation gives $|\,\chi(\bar{t},r)|$ $\neq$ $0$,
so that the background fields $\lambda(\bar{t},r)$ and ${\cal R}(\bar{t},r)$ are
smooth. This implies that the solution $\Theta(\bar{t},r)$ $\in$
$S_N(\bar{t})$, with $N\geq 1$, is continuous  and
must cross the value $+\,\pi/4$ at least once.
Define $r_+$ to be the largest radial distance for which
$\Theta(\bar{t},r)$ $=$ $+\,\pi/4$.
In order for $\Theta(\bar{t},r)$ to reach the asymptotic value $N\,\pi$, the slope
of $\Theta(\bar{t},r)$ at $r=r_+$ clearly must be nonnegative,
\beq \label{ThetaSN}
\partial_r \Theta(\bar{t},r_+)\geq 0\;,
\eeq
as long as $\Theta$ $\in$ $S_N(\bar{t})$, with $N\geq 1$.

On the other hand,
Eqs. (\ref{ineq:lambdaMR}) and (\ref{FZMeqtlarge}), together with the fact
that $\sin 2\Theta (\bar{t},r_+)=+1$, give the following inequality:
\beq\label{Thetabound}
\partial_r \Theta(\bar{t},r_+)=-\,\lambda(\bar{t},r_+)+{\cal
R}(\bar{t},r_+)<0\;.
\eeq
This last result contradicts the earlier result (\ref{ThetaSN}), which was
based on the assumption that
$\Theta (\bar{t},r)\in S_{N}(\bar{t})$, with $N\geq 1$.
Hence, $\Theta(\bar{t},r)$ $\notin$ $S_N(\bar{t})$, for positive integer $N$.
The case of \emph{negative} integer $N$ is ruled out in the same way.
The conclusion is thus that $\Theta (\bar{t},r)$ belongs to $S_0(\bar{t})$.

For strongly dissipative spherically symmetric $SU(2)$ gauge field
solutions, we find that the rotator ${\cal R}(t,r)$ at large times
lacks the strength to overcome the resistance of the deviator ${\cal
D}(t,r)$, so as to give
a nonzero spinor twist number at large times. In short, we have
\beq
\lim_{t\rightarrow -\infty}N_{\Theta}(t)=
\lim_{t\rightarrow +\infty}N_{\Theta}(t)=0.
\label{Nplusminus-big-t}
\eeq
This result shows that the twist factor  $\Delta N_{\Theta}[ +\infty, -\infty \,]$
$\equiv$ $N_{\Theta}(+\infty) - N_{\Theta}(-\infty)$ $=$ $0$ does not contribute to
the spectral flow (\ref{SelectionRule}), at least for the case of
strongly dissipative spherically symmetric $SU(2)$ gauge field solutions.

\end{appendix}

\end{document}